\documentclass[11pt]{article}
\usepackage{graphicx}
\usepackage{graphicx}
\usepackage{dcolumn}
\usepackage{amsmath}
\usepackage{epsfig}
\usepackage{multirow}
\usepackage{relsize}
\RequirePackage{xspace}

\def\Bztorhozrhoz {\ensuremath{\Bz \to \rho^0 \rho^0 }\xspace}
\def\Btozz {\ensuremath{\Bz \to \rho^0 \rho^0 }\xspace}

\def\Bztorhoprhom {\ensuremath{\Bz \to \rho^+ \rho^- }\xspace}
\def\Bptorhozrrhop {\ensuremath{\Bp \to \rho^+ \rho^0 }\xspace}

\def\babar{\mbox{\slshape B\kern-0.1em{\smaller A}\kern-0.1em B\kern-0.1em{\smaller A\kern-0.2em R}}}
\def\Bbar    {\kern 0.18em\overline{\kern -0.18em B}{}\xspace}
\def\Dbar    {\kern 0.18em\overline{\kern -0.18em D}{}\xspace}
\def\Kbar    {\kern 0.18em\overline{\kern -0.18em K}{}\xspace}
\def\pep2{PEP-II}
\mathchardef\Upsilon="7107
\newcommand{\optbar}[1]{\shortstack{{\tiny (\rule[.4ex]{1em}{.1mm})}
  \\ [-.7ex] $#1$}}
\def\BorBbar    {\kern 0.18em\optbar{\kern -0.18em B}{}\xspace}
\def\DorDbar    {\kern 0.18em\optbar{\kern -0.18em D}{}\xspace}
\def\KorKbar    {\kern 0.18em\optbar{\kern -0.18em K}{}\xspace}

\def\qqbar {\ensuremath{q\overline q}\xspace}

\def\BB      {\ensuremath{B\Bbar}\xspace}
\def\Bz      {\ensuremath{B^0}\xspace}
\def\Bzb     {\ensuremath{\Bbar^0}\xspace}
\def\BzBzb   {\ensuremath{\Bz {\kern -0.16em \Bzb}}\xspace}
\def\Bu      {\ensuremath{B^+}\xspace}
\def\Bub     {\ensuremath{B^-}\xspace}
\def\Bp      {\ensuremath{\Bu}\xspace}

\def\BpBm    {\ensuremath{\Bu {\kern -0.16em \Bub}}\xspace}

\def\CP                {\ensuremath{C\!P}\xspace}
\def\pep2{PEP-II}
\mathchardef\Upsilon="7107
\def\Y#1S{\ensuremath{\Upsilon{(#1S)}}\xspace}

\def\FourS {\Y4S}

\def\BR         {{\ensuremath{\cal B}\xspace}}
\def\Bztorhozrhoz {\ensuremath{\,\Bz \to \rho^0\rho^0}\xspace}
\def\Bztorhoprhom {\ensuremath{\,\Bz \to \rho^+\rho^-}\xspace}

\def\Bztorhozfz {\ensuremath{\,\Bz \to \rho^0 f_0(980)}\xspace}
\def\Bztofzfz   {\ensuremath{\,\Bz \to f_0(980) f_0(980)}\xspace}

\def\DeltaE {\ensuremath{\Delta E}\xspace}
\def\mes{\ensuremath{m_{\mathrm{ES}}}\xspace}

\def\eshape{\ensuremath{{\cal E}\text{-shape}}\xspace}

\newcommand{\tev}{\ensuremath{\mathrm{\,Te\kern -0.1em V}}\xspace}
\newcommand{\gev}{\ensuremath{\mathrm{\,Ge\kern -0.1em V}}\xspace}
\newcommand{\mev}{\ensuremath{\mathrm{\,Me\kern -0.1em V}}\xspace}
\newcommand{\kev}{\ensuremath{\mathrm{\,ke\kern -0.1em V}}\xspace}
\newcommand{\ev}{\ensuremath{\mathrm{\,e\kern -0.1em V}}\xspace}
\newcommand{\gevc}{\ensuremath{{\mathrm{\,Ge\kern -0.1em V\!/}c}}\xspace}
\newcommand{\mevc}{\ensuremath{{\mathrm{\,Me\kern -0.1em V\!/}c}}\xspace}
\newcommand{\gevcc}{\ensuremath{{\mathrm{\,Ge\kern -0.1em V\!/}c^2}}\xspace}
\newcommand{\mevcc}{\ensuremath{{\mathrm{\,Me\kern -0.1em V\!/}c^2}}\xspace}

\newcommand{\jprlBase}       {Phys.\ Rev.\ Lett.\xspace}
\newcommand{\jprl}      [1]  {\jprlBase\ {\bf #1}}

\newcommand{\jprBase}        {Phys.\ Rev.\xspace}
\newcommand{\jprd}      [1]  {\jprBase\ D~{\bf #1}}

\def\qqbar {\ensuremath{q\overline q}\xspace}
\def\u     {\ensuremath{u}\xspace}

\def\d     {\ensuremath{d}\xspace}

\def\b     {\ensuremath{b}\xspace}

\def\pip   {\ensuremath{\pi^+}\xspace}
\def\pim   {\ensuremath{\pi^-}\xspace}

\def\B       {\ensuremath{B}\xspace}
\def\mes        {\mbox{$m_{\rm ES}$}\xspace}
\def\DeltaE     {\mbox{$\Delta E$}\xspace}

\def\epem       {\ensuremath{e^+e^-}\xspace}

\newcommand{\eg}{{\em e.g.}}

\newcommand{\BABARPubYear}    {07}
\newcommand{\BABARConfNumber} {012}
\newcommand{\SLACPubNumber} {12734}
\newcommand{\LANLNumber} {0708.1630}

\long\def\inst#1{\par\nobreak\kern 4pt\nobreak
    {\it #1}\par\vskip 10pt plus 3pt minus 3pt}

\begin{document}
{\pagestyle{empty}

\begin{flushright}
\babar-CONF-\BABARPubYear/\BABARConfNumber \\
SLAC-PUB-\SLACPubNumber \\
arXiv:\LANLNumber [hep-ex]\\
August 2007 \\
\end{flushright}

\par\vskip 5cm

\begin{center}
\Large \bf Time-Dependent Analysis of the Decay $B^0\to\rho^0\rho^0$
\end{center}
\bigskip

\begin{center}
\large The \babar\ Collaboration\\
\mbox{ }\\
August 12, 2007
\end{center}
\bigskip \bigskip

\begin{center}
\large \bf Abstract
\end{center}

We study the decay \Btozz in a sample of about 427 million
$\Upsilon (4S)\rightarrow B\kern 0.18em\overline{\kern -0.18em B}$
decays collected with the $\babar$ detector at the
PEP-II asymmetric-energy $e^+e^-$ collider at SLAC.
We find the branching fraction 
$\BR = (0.84\pm 0.29\pm 0.17)\times 10^{-6}$
and longitudinal polarization fraction of
$f_L = 0.70\pm 0.14\pm 0.05$, where the first uncertainty is
statistical, and the second is systematic.
The evidence for the \Btozz\ signal has $3.6\sigma$ significance.
We investigate the proper-time dependence of the 
longitudinal component in the decay and measure the
$C\!P$-violating coefficients 
$S^{00}_L = (0.5\pm0.9\pm0.2)$
and $C^{00}_L = (0.4\pm0.9\pm0.2)$,
corresponding to the sine and cosine terms in the
time evolution of asymmetry.
We study the implication of these results for 
penguin contributions in $B\to\rho\rho$ decays
and for the CKM unitarity angle $\alpha$.

\vfill
\begin{center}
Contributed to the 
XXIII$^{\rm rd}$ International Symposium on Lepton and Photon Interactions at High~Energies, 8/13 -- 8/18/2007, Daegu, Korea
\end{center}

\vspace{1.0cm}
\begin{center}
{\em Stanford Linear Accelerator Center, Stanford University, 
Stanford, CA 94309} \\ \vspace{0.1cm}\hrule\vspace{0.1cm}
Work supported in part by Department of Energy contract DE-AC03-76SF00515.
\end{center}

\newpage
} 

%
%
\begin{center}
\small

The \babar\ Collaboration,
\bigskip

%
{B.~Aubert,}
{M.~Bona,}
{D.~Boutigny,}
{Y.~Karyotakis,}
{J.~P.~Lees,}
{V.~Poireau,}
{X.~Prudent,}
{V.~Tisserand,}
{A.~Zghiche}
\inst{Laboratoire de Physique des Particules, IN2P3/CNRS et Universit\'e de Savoie, F-74941 Annecy-Le-Vieux, France }
{J.~Garra~Tico,}
{E.~Grauges}
\inst{Universitat de Barcelona, Facultat de Fisica, Departament ECM, E-08028 Barcelona, Spain }
% ...

\end{center}\newpage


\section{INTRODUCTION}
\label{sec:Introduction}

Measurements of \CP-violating asymmetries in the \BzBzb system 
test the flavor structure of the standard model by 
over-constraining the Cabibbo-Kobayashi-Maskawa (CKM)
quark-mixing matrix~\cite{CabibboKobayashi}.
The time-dependent \CP asymmetry in the decays of \Bz\ or \Bzb\ mesons
to a \CP eigenstate dominated by the tree-level amplitude
$\b \to \u{\bar\u}\d$
measures $\sin 2\alpha_\mathrm{eff}$, where
$\alpha_\mathrm{eff}$ differs from the CKM unitarity
triangle angle $\alpha\equiv
\arg\left[-V_{td}^{}V_{tb}^{*}/V_{ud}^{}V_{ub}^{*}\right]$ by a
quantity  $\Delta\alpha$ accounting for the contributions from 
loop (penguin) amplitudes.
The value of $\Delta\alpha$ can be extracted from an analysis 
of the branching fractions of the $B$ decays into the full 
set of isospin-related channels~\cite{gronau90}.

Branching fractions and time-dependent \CP
asymmetries in $B\to\pi\pi$, $\rho\pi$, and $\rho\rho$
have already provided information on $\alpha$.
Since the tree contribution to the $B^0\to\rho^0\rho^0$~\cite{footnote}
decay is color-suppressed,
the decay rate is sensitive to the penguin amplitude.
The $\Bz\to\rho^0\rho^0$ decay has a much smaller branching 
fraction than $\Bz\to\rho^{+}\rho^{-}$ and $B^{+}\to\rho^{+}\rho^0$
channels~\cite{vvbabar,rho0rhopbelle,rho0rhop2,rhoprhom,rhoprhombelle},
which leads to a more stringent limit on
$\Delta\alpha$ from isospin
analysis~\cite{gronau90,rhoprhom,falketal} than is possible  
in $\pi\pi$ system. 
This makes the $\rho\rho$ system particularly effective for
measuring~$\alpha$.

The error due to the penguin contribution becomes
the dominant uncertainty in the measurement of $\alpha$ using
$B\to\rho\rho$ decays. However, in contrast to $B\to\pi^0\pi^0$ decays,
the four-track final state makes a
time-dependent analysis of $B^0\to\rho^0\rho^0$ decays
feasible. It allows us
to measure the \CP\ parameters $S^{00}_{L}$ and $C^{00}_{L}$ directly,
analogous to $S^{+-}_{L}$ and $C^{+-}_{L}$,
resolving ambiguities inherent to isospin triangle orientations.
The $C^{00}$ coefficient is associated with
the difference in decay amplitudes for 
$B \to \rho^0\rho^0$ and $\Bbar \to \rho^0\rho^0$,
while the $S^{00}$ coefficient involves interference between the 
$\Bz - \Bzb$ mixing and decay amplitudes.

In $B\to\rho\rho$ decays the final state is
a superposition of \CP-odd and \CP-even states. 
An isospin-triangle relation~\cite{gronau90} holds for each
of the three helicity amplitudes, which can be separated through
an angular analysis. 
The helicity angles $\theta_1$ and $\theta_2$ are defined 
as the angles between the direction of $\pi^+$ and the direction 
of the \B in the rest system of each of the $\rho^0$ candidates.
The resulting angular distribution
is given by
\begin{eqnarray}
{d^2\Gamma / (\Gamma\,d\!\cos \theta_1\,d\!\cos \theta_2)}=
\frac{9}{4} \left \{ \frac{1}{4} (1 - f_L)
\sin^2 \theta_1 \sin^2 \theta_2 + f_L \cos^2 \theta_1 \cos^2 \theta_2 \right\},
\label{eq:helicityshort}
\end{eqnarray}
\noindent where $f_L=|A_0|^2/(\Sigma|A_\lambda|^2)$ is the
longitudinal polarization fraction and
$A_{\lambda=-1,0,+1}$ are the helicity amplitudes.
The fraction of longitudinal polarization is {\it a priori} unknown.
Polarization was expected to be predominantly 
longitudinal~\cite{polarizationtheory}. However, significant
departure from this expectation was found in penguin-dominated 
$B$-decay modes~\cite{vv} and polarization and
$C\!P$-asymmetry measurements in the
$B^0\to\rho^0\rho^0$ decay may help in resolving 
this puzzle~\cite{polarizationrho0rho0}.

In this paper, we update our previous measurement of the branching
fraction and  longitudinal polarization fraction in
$B^0\to\rho^0\rho^0$ decays~\cite{vvbabar}, and present the first
study of the time-dependent $C\! P$ asymmetry $\mathcal{A}_{C\!P}$ in
this mode. 
We determine the coefficients $C_L^{00}$ and $S_L^{00}$ of the
asymmetry for the longitudinal component, which is given by
\begin{equation}
\mathcal{A}_{C\!P}(\Delta t)=-C^{00}_L \cos{\Delta m\Delta t}+S^{00}_L \sin{\Delta m\Delta t}
\end{equation}
These coefficients, together with improved measurements
of the branching fraction and longitudinal polarization,
allow a complete isospin analysis and improved constraints
on the penguin contribution to $B\to\rho\rho$ decays.
Changes with respect to our previous analysis~\cite{vvbabar} 
include increased datasample, improved track-selection techniques,
and inclusion of the $B$-decay time information.


\section{DETECTOR AND DATASET}
\label{sec:Detector}

These results are based on data collected
with the \babar\ detector~\cite{babar} at the PEP-II asymmetric-energy
$e^+e^-$ collider~\cite{pep2}.
A sample of $427\pm 5$ million $\BB$ pairs
was recorded at the $\FourS$ resonance with the center-of-mass 
(c.m.) energy $\sqrt{s} = 10.58$~\gev.
Charged-particle momenta and trajectories are measured in a tracking system
consisting of a five-layer double-sided silicon vertex tracker
and a 40-layer drift chamber,
both within a 1.5-T solenoidal magnetic field.
Charged-particle identification is provided by
measurements of the energy loss
in the tracking devices and by a ring-imaging Cherenkov detector.


\section{ANALYSIS METHOD}
\label{sec:analysis}

We select $\B\to M_1M_2\to(\pi^+\pi^-)(\pi^+\pi^-)$
candidates, with $M_{1,2}$ standing for a $\rho^0$ or $f_0$ candidate,
from neutral combinations of four charged tracks that
are consistent with originating from a single vertex near
the $e^+e^-$ interaction point. We veto tracks that are 
identified as kaons or electrons.
The identification of signal $B$ candidates is based
on several kinematic variables. 
The beam-energy-substituted mass,
$\mes = [(s/2 + {\mathbf {p}}_i\cdot {\mathbf{p}}_B)^2/E_i^2-
{\mathbf {p}}_B^2]^{1/2}$,
where the initial $e^+e^-$
four-momentum $(E_i, {\mathbf {p_i}})$ and the \B
momentum ${\mathbf {p_B}}$ are defined in the laboratory frame, is
centered near the \B mass with a resolution of $2.6~\mevcc$ for signal
candidates.  
The difference $\DeltaE = E_B^{\rm cm} - \sqrt{s}/2$ between the
reconstructed \B energy in the 
c.m. frame and its known value $\sqrt{s}/2$ has a maximum near zero with a
resolution of $20~\mev$ for signal events. Four other kinematic
variables describe two possible
$\pi^+\pi^-$ pairs: the invariant masses $m_{1}$, $m_{2}$
and the helicity angles $\theta_1,\ \theta_2$. 

The selection requirements for signal candidates are the following:
$5.245 < \mes < 5.290~\gevcc$, 
$|\DeltaE|<$ 85~\mev,
$550< m_{1,2} < 1050~\mevcc$,
and $|\cos\theta_{1,2}|<0.98$.
The last requirement removes a region corresponding to low-momentum 
pions with low and uncertain reconstruction efficiency.
In addition, we veto the copious decays 
$\Bz\to D^{(*)-}\pip\to(h^+\pim\pim)\pip$, 
where $h^+$ refers to a pion or kaon, by requiring the
invariant mass of the three-particle combination
to differ from the $D$-meson mass by more 
than $13.2~\mevcc$, or $40~\mevcc$ if one of the particles is consistent with a
kaon hypothesis. 

We reject the dominant  $\epem\to q\bar{q}\ (q=u,d,s,c)$ (continuum)
background by requiring $|\cos\theta_T| < 0.8$, where $\theta_T$
is the angle between the $B$-candidate thrust axis
and that of the remaining tracks and neutral clusters in
the event, calculated in the c.m. frame.
We further suppress continuum background using 
a neural network-based discriminant $\mathcal{E}$, which combines eight
topological variables calculated in the c.m. frame.
In addition to $\cos\theta_T$,
they are the polar angles of the $B$ momentum vector 
and the $B$-candidate thrust axis with respect to the beam axis, 
the value of the event thrust,
two Legendre moments $L_0$ and $L_2$ of the energy
flow around the $B$-candidate thrust axis~\cite{bigPRD} computed
separately for neutral and charged particles, 
and the sum of the transverse momenta of all particles in the rest
of the event, calculated with respect to the $B$ direction.

We use multivariate $B$-flavor-tagging
algorithms trained to identify primary leptons, kaons, soft pions,
and high-momentum charged particles
from the other $B$, called $B_{\rm tag}$~\cite{babarsin2beta}.
The effective tagging efficiency, which takes into account the
efficiency to find the tag and the mistag probability, is $(31.1\pm
0.3)\%$, as determined on a sample of fully
reconstructed open-charm decays. We use both tagged and untagged
events in our sample. 
Additional background discrimination power arises from the difference 
between the tagging efficiencies for signal and background in seven
tagging categories ($c_{\rm tag}=1..7$).
We determine the proper time difference $\Delta t$ between 
the signal $B$ and $B_{\rm tag}$ from the spatial separation between 
their decay vertices. The $B_{\rm tag}$ vertex is reconstructed from the
remaining charged tracks in the event and its uncertainty dominates
the $\Delta t$ resolution $\sigma_{\Delta t}$. The average proper time
resolution is $\langle\sigma_{\Delta t}\rangle \approx 0.7$~ps. Only
events that satisfy $|\Delta t|<15$~ps and 
$\sigma_{\Delta t}<2.5$~ps are retained. 

After application of all selection criteria,
$N_{\rm cand}=65637$ events form the sample for the maximum likelihood
fit.
On average, each selected event has $1.05$ signal candidates, 
while in Monte Carlo~\cite{GEANT}
samples of longitudinally and transversely
polarized $B^0\to\rho^0\rho^0$ decays
we find $1.15$ and $1.03$ candidates, respectively.
When more than one candidate is present in the same event,
the candidate having the best $\chi^2$ consistency
with a single four-pion vertex is selected. 
Simulation shows that 18\% of longitudinally
and 4\% of transversely polarized $\Btozz$
events are misreconstructed with one or more tracks
not originating from the $B^0\to\rho^0\rho^0$ decay.
These are mostly due to combinatorial background from
low-momentum tracks from the other \B meson in the event. Such events
still carry the characteristic topology of $B^0\to\rho^0\rho^0$
events; they are modeled separately from the perfectly reconstructed
events and included into the probability density functions.


\section{MAXIMUM LIKELIHOOD FIT}
\label{sec:fit}

We use an unbinned extended maximum likelihood fit to extract
the $B^0\to\rho^0\rho^0$ event yield and fraction of longitudinal
polarization $f_L$. We also fit for the event yields of $B^0\to\rho^0f_0$ 
and $B^0\to f_0f_0$ decays, as well as yields in several background categories.
The likelihood function is
\begin{equation}
{\cal L} = \exp\left(-\sum_{k}^{} n_{k}\right)\,
\prod_{i=1}^{N_{\rm cand}}
\left(\sum_{j}~n_{j}\,
{\cal P}_{j}(\vec{x}_{i})\right),
\label{eq:likel}
\end{equation}
where $n_j$ is the unconstrained (except if noted otherwise)
number of events for each event type $j$
($B^0\to\rho^0\rho^0$ , $B^0\to\rho^0f_0(980)$, $B^0\to f_0(980)f_0(980)$,
several background components from exclusive and inclusive \B decays, 
and continuum), and ${\cal P}_{j}(\vec{x}_{i})$ is the 
probability density function (PDF) of the variables
$\vec{x}_{i}=\{m_{\rm{ES}}, \Delta E, \mathcal{E},
m_1, m_2, \cos\theta_1, \cos\theta_2, c_{\rm tag},
\Delta t, \sigma_{\Delta t} \}_i$
for the $i$th event.

Since the statistical correlations among most variables are found to be small,
we take each ${\cal P}_j$ as the product of the PDFs for the
separate variables. In a number of special cases, for the 
mass-helicity PDF of continuum backgrounds, or for the mass and
helicity PDFs of the \B backgrounds, we use 2- or 4-dimensional PDFs
to properly describe the  kinematic correlation between the
observables. 

We use double-Gaussian functions 
to parameterize the \mes\ PDFs for fully-reconstructed signal events,
double-Gaussian functions for  $\Delta E$ 
and relativistic Breit-Wigner functions 
for the resonance masses of $\rho^0$
and $f_0(980)$~\cite{f0mass}.
The angular distribution at production for 
$B^0\to\rho^0\rho^0$, $B^0\to \rho^0f_0$, and $B^0\to f_0f_0$
modes (expressed as a function of the longitudinal 
polarization in Eq.~(\ref{eq:helicityshort}) for 
\Btozz) is multiplied by a detector acceptance function 
${\cal G}(\cos\theta_1, \cos\theta_2)$,
determined from Monte Carlo. 
The distributions of misreconstructed signal events
are parameterized with empirical shapes in a way similar
to that used for $B$ background discussed below.
The neural network discriminant ${\cal E}$ 
is described by two (continuum) or three (\B-decay events) asymmetric
Gaussian functions with different parameters for signal
and background distributions. 

The PDFs for inclusive \B decay modes are
generally modeled with empirical analytical distributions.
Several variables have distributions
similar or identical to those for signal, such as $m_{\rm{ES}}$
when all four tracks come from the same $B$, or $\pi^+\pi^-$
invariant mass $m_{1,2}$ when both tracks come from
a $\rho^0$ meson.
Also for some of the modes the two $\pi^+\pi^-$ pairs
can have different mass and helicity distributions, 
\eg\ when only one of the two combinations 
comes from a genuine $\rho^0$ or $f_0$ meson, 
or when one of the two pairs contains a
high-momentum pion (as in $B\to a_1\pi$). In such cases,
we use a four-dimensional mass-helicity PDF.

The proper-time distribution for signal and background \B decays 
\begin{equation}
f(\Delta t,Q)\sim \frac{e^{-\left|\Delta t\right|/\tau}}{4\tau} 
\times \bigg\{ 1 - Q \Delta w + Q\mu(1-2\omega) +\left[Q(1-2w)+\mu
(1-Q\Delta \omega)\right]\mathcal{A}_{C\!P}(\Delta t)\bigg\}
\end{equation}
is convolved with a resolution function, modeled as a superposition
of three Gaussian distributions with the means and widths scaled by the
per-event error on $\Delta t$. Here $Q$ is the flavor of
$B_\mathrm{tag}$, $w$ is 
the average mistag probability, and $\Delta w$ and $\mu$ parameters
describe the difference in mistag probability and the tagging efficiency
asymmetry between $B^0$ and $\overline{B}^0$ mesons. The time
distribution of continuum 
background is assumed to have zero lifetime. 

The signal and $B$-background PDF parameters are extracted from
simulation. The Monte Carlo parameters for
$m_{\rm{ES}}$, $\Delta E$, and ${\cal E}$ PDFs are adjusted by
comparing data and simulation in control channels with similar
kinematics and topology,
such as $B^0\to D^-\pi^+$ with $D^-\to K^+\pi^-\pi^-$.
The continuum background PDF shapes are extracted 
from off-resonance data or on-resonance sideband data, with parameters
of the most discriminating PDFs ($m_\mathrm{ES}$, $\Delta E$, ${\cal
  E}$) left free in the final fit. 
The discrete $B$-flavor tagging PDFs and parameters of the proper time
distributions for signal modes are 
obtained in dedicated fits to events with identified exclusive \B
decays~\cite{babarsin2beta}. 
The tagging PDFs for inclusive \B backgrounds are determined
by Monte Carlo and their systematic uncertainties are studied in data. 

Backgrounds from ``charmless'' $b\to u$ transitions, in particular
events containing $\rho$, $f_0$, or $K^{*}$ mesons, have kinematic
distributions similar to those of signal events. We study the
contributions of the dominant decay modes in high-statistics exclusive
Monte Carlo samples. We also develop two complementary strategies to
model these backgrounds in the likelihood fit. 

In the first approach, we single out contributions from the following
dominant modes: $B^0\to a_1^{\pm}\pi^{\mp}$, $B^0\to \rho^0K^{*0}$, $B^0\to
f_0K^{*0}$, $B^+\to\rho^+\rho^0$,
$B^+\to a_1^{0}\pi^{+}$, $B\to\eta'K$, and $B^+\to\rho^0\pi^+$. The contribution of the 
$B^0\to a_1^{\pm}\pi^{\mp}$ decays includes both events where all four
tracks are correctly associated to the \B candidate, and events where
at least one of the tracks is picked from the other \B meson
(so-called  $B^0\to a_1^{\pm}\pi^{\mp}$ 
{\em ``self-crossfeed''\/} events). The
event yield of the $B^0\to a_1^{\pm}\pi^{\mp}$ decays is allowed to
vary in the fit, while the yields of other six charmless modes listed
above are fixed to the expected values~\cite{PDG2006, HFAG07, a10pi}. The
events from open charm $b\to c$ transitions are parameterized as a
separate background component, with its yield allowed to vary in the
data fit. 

In the second strategy, we split the \B background into three distinct
categories: $B^0\to a_1^{\pm}\pi^{\mp}$ decays where all four
tracks are correctly associated with the \B candidate, an appropriately
weighted combination of dominant charmless decays, and the rest of the
generic $b\to u$, $b\to c$, $b\to s$, and $b\to d$ transitions. For
the charmless event category, we combine the following modes:  
$B^0\to a_1^{\pm}\pi^{\mp}$ self-crossfeed events, 
$B^0\to \rho^0K^{*0}$, $B^0\to f_0K^{*0}$, $B^0\to\rho^+\rho^-$,
$B^0\to \rho^{\pm}\pi^{\mp}$, 
$B^+\to\rho^+\rho^0$, 
$B^+\to a_1^{0}\pi^{+}$, $B^+\to a_1^{+}f_0$, $B\to\eta'K$, and
$B^+\to\rho^0\pi^+$. 
Kinematic distributions in these events, especially events in which at
least one charged particle was not correctly associated to the \B
candidate, are similar to each other, and
also to other, poorly measured charmless decays. This allows us to
vary the overall event yield for this category of events, after fixing
the relative weights of each mode to the expected
values~\cite{PDG2006, HFAG07, a10pi}. For yet-unmeasured 
$\BR(B^+\to a_1^{+}f_0)$, we assume a conservative value of 
$(1\pm1)\times 10^{-5}$; this branching ratio corresponds to the
expectation of $10\pm10$ events in the selected data sample. Event
yields associated with  
$B^0\to a_1^{\pm}\pi^{\mp}$ component and the generic \B decays are
allowed to vary in the maximum likelihood fit.

We find that both strategies for describing \B decay backgrounds
presented above adequately describe the data, and are in excellent
agreement on the 
yields, polarization, and $C\! P$ parameters of $B^0\to\rho^0\rho^0$
decays. Statistical correlation between the two models is
high (93\% for the yields and $f_L$, 85\% for $S_L^{00}$ and
87\% for $C_L^{00}$, as determined from a number of Monte Carlo
experiments with event composition matched to the data). However, the
two approaches differ in their  
sensitivity to the variations of the background composition, and have
mostly uncorrelated systematic uncertainties associated with the \B
background PDFs and fit bias. 
Our two models
represent the extreme approaches to describing the \B
backgrounds: one relies on the knowledge of the branching ratios
associated with the dominant \B backgrounds, and the other relies on the
modeling of the kinematic distributions of \B backgrounds in general. 
To further reduce these systematic effects, we average the results of
the two models after the maximum likelihood fits. 

Other four-pion final states,
such as $B^0\to \rho^0\pi^+\pi^-$ and
$B^0\to\pi^+\pi^-\pi^+\pi^-$, require special care. The rates for
these modes are not well constrained~\cite{ref:Belle_a1pi}, although
their contributions are expected to 
be small in our invariant mass window.
We parameterize the PDFs associated with these modes using the exclusive
Monte Carlo samples, which assume uniform phase-space distributions of
the final state mesons, and allow their yields to vary in the fit to
the data. The absolute maximum of the likelihood occurs in the unphysical
(negative) region for these yields, and we restrict the branching
ratios for each of these modes to the range $[0..6]\times 10^{-6}$. 
The upper limit corresponds to several times the rate measured in
Ref.~\cite{ref:Belle_a1pi}. With these restrictions, the yields for
both $B^0\to \rho^0\pi^+\pi^-$ and $B^0\to\pi^+\pi^-\pi^+\pi^-$ decays 
converge identically to zero in the maximum likelihood fit, indicating
no evidence for any contribution from these modes to the fit
region.


\section{RESULTS}
\label{sec:results}

\begin{figure}[h!]
\begin{center}
\begin{tabular}{cc}
\includegraphics[height=1.5in]{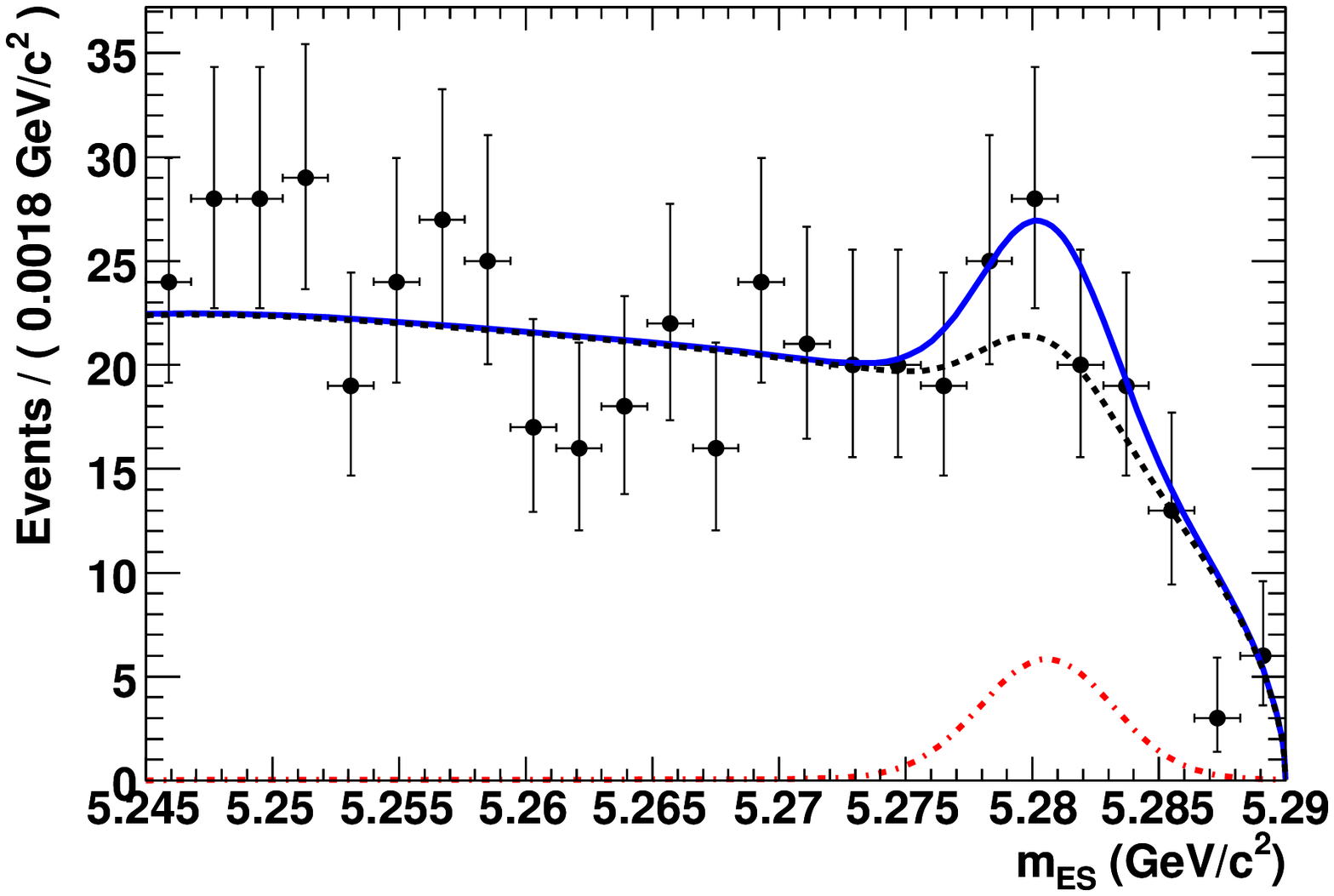}
&
\includegraphics[height=1.5in]{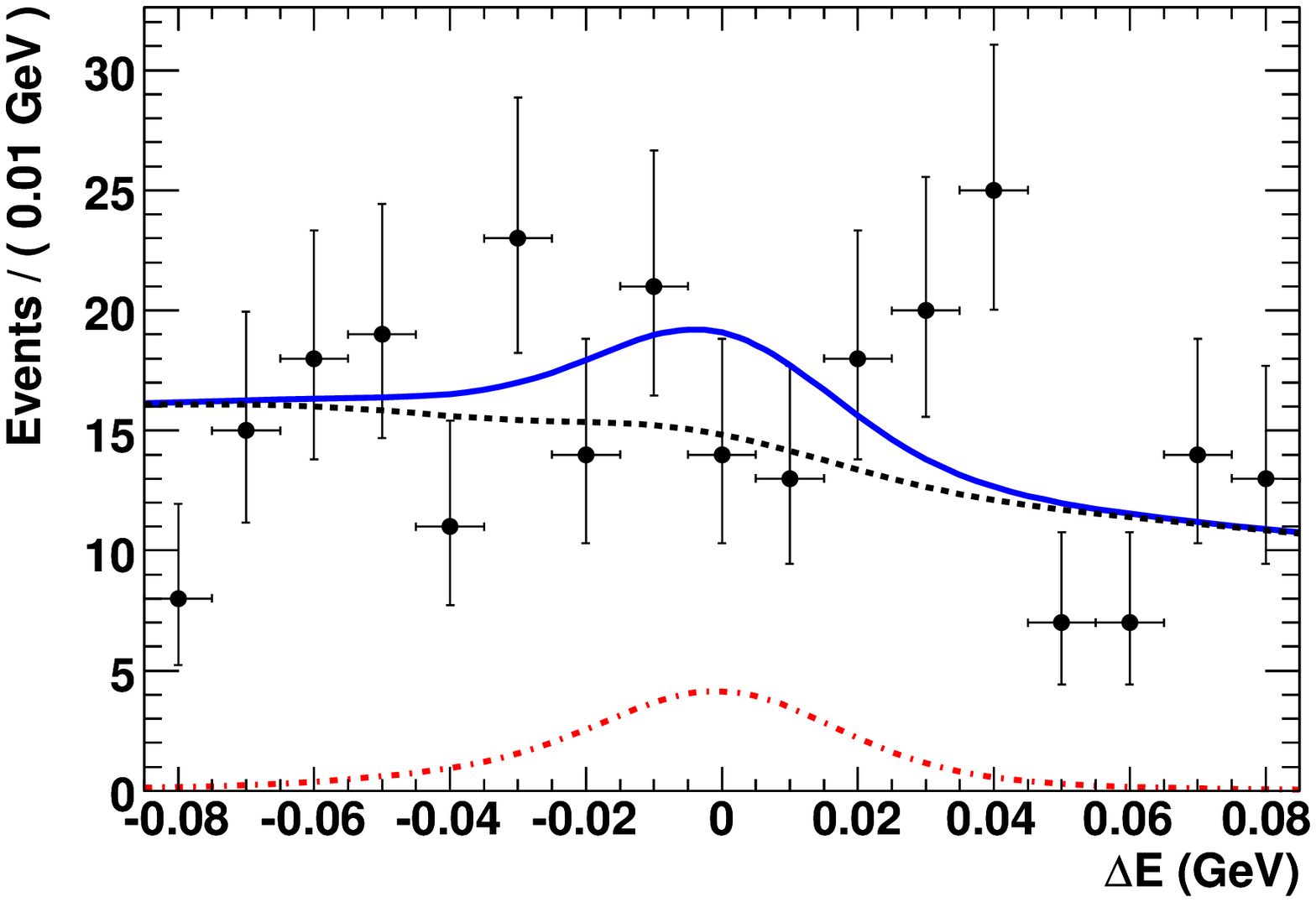}
\\
(a) & (b) \\
\includegraphics[height=1.5in]{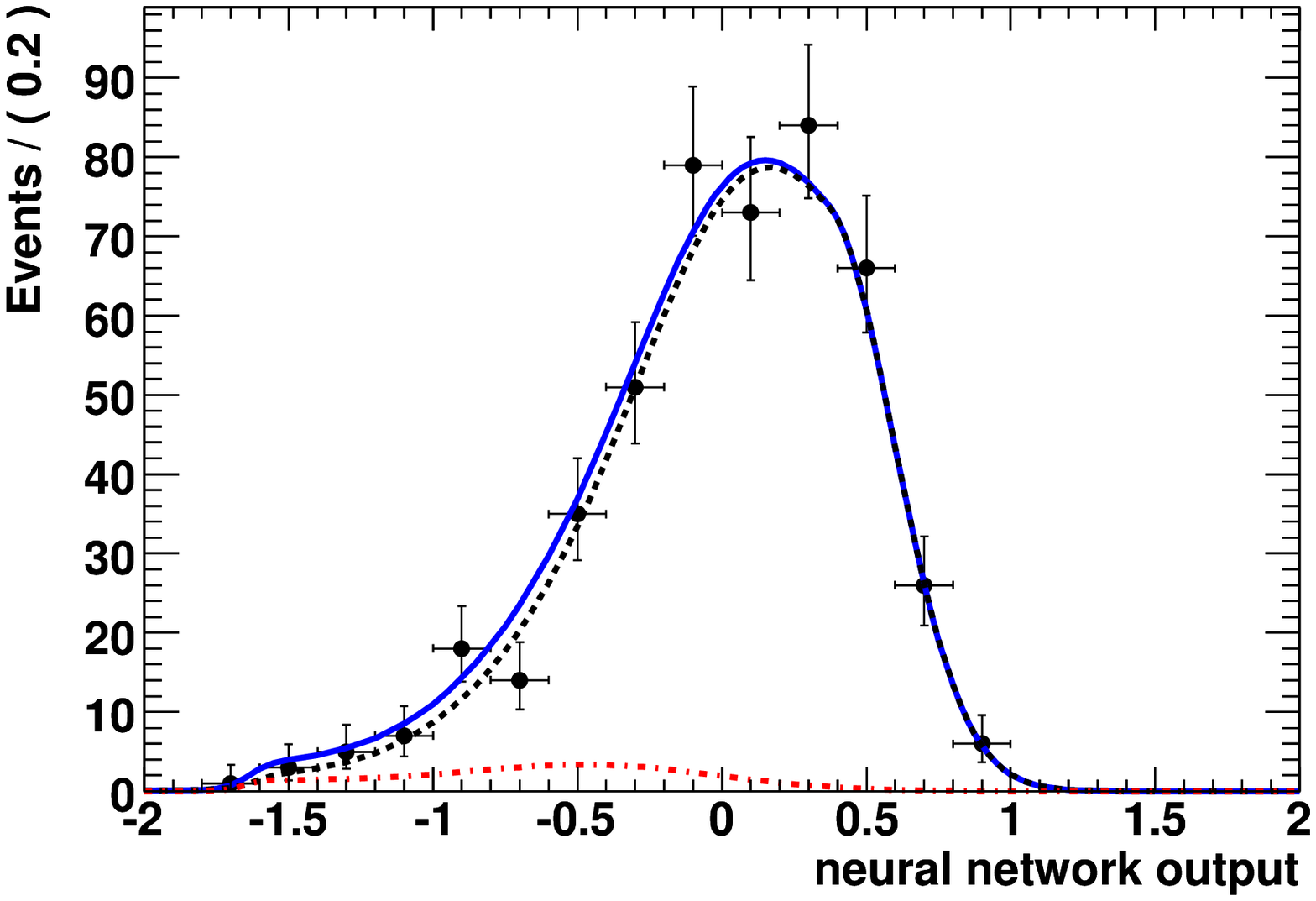}
&
\\
(c) & \\
\includegraphics[height=1.5in]{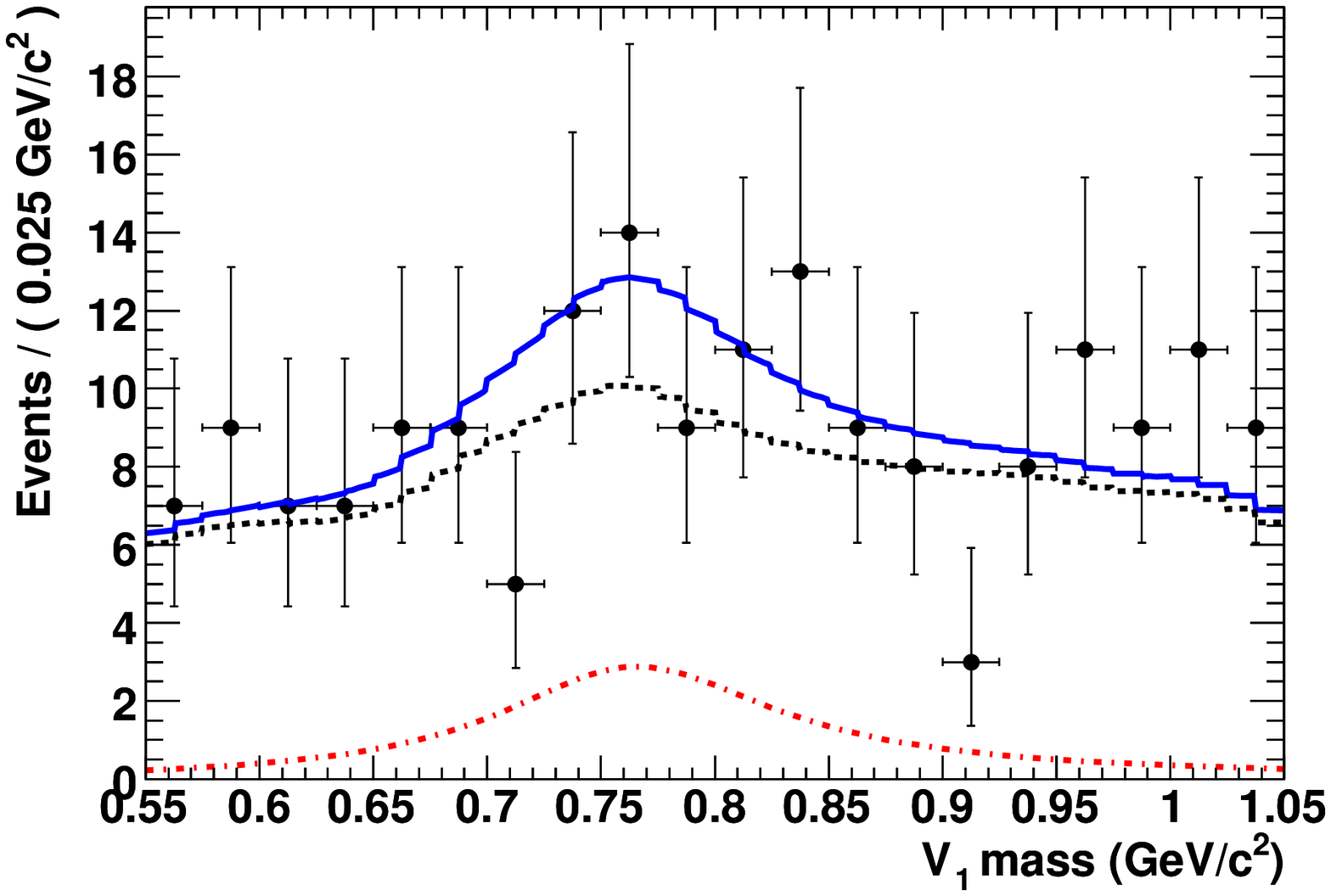}
&
\includegraphics[height=1.5in]{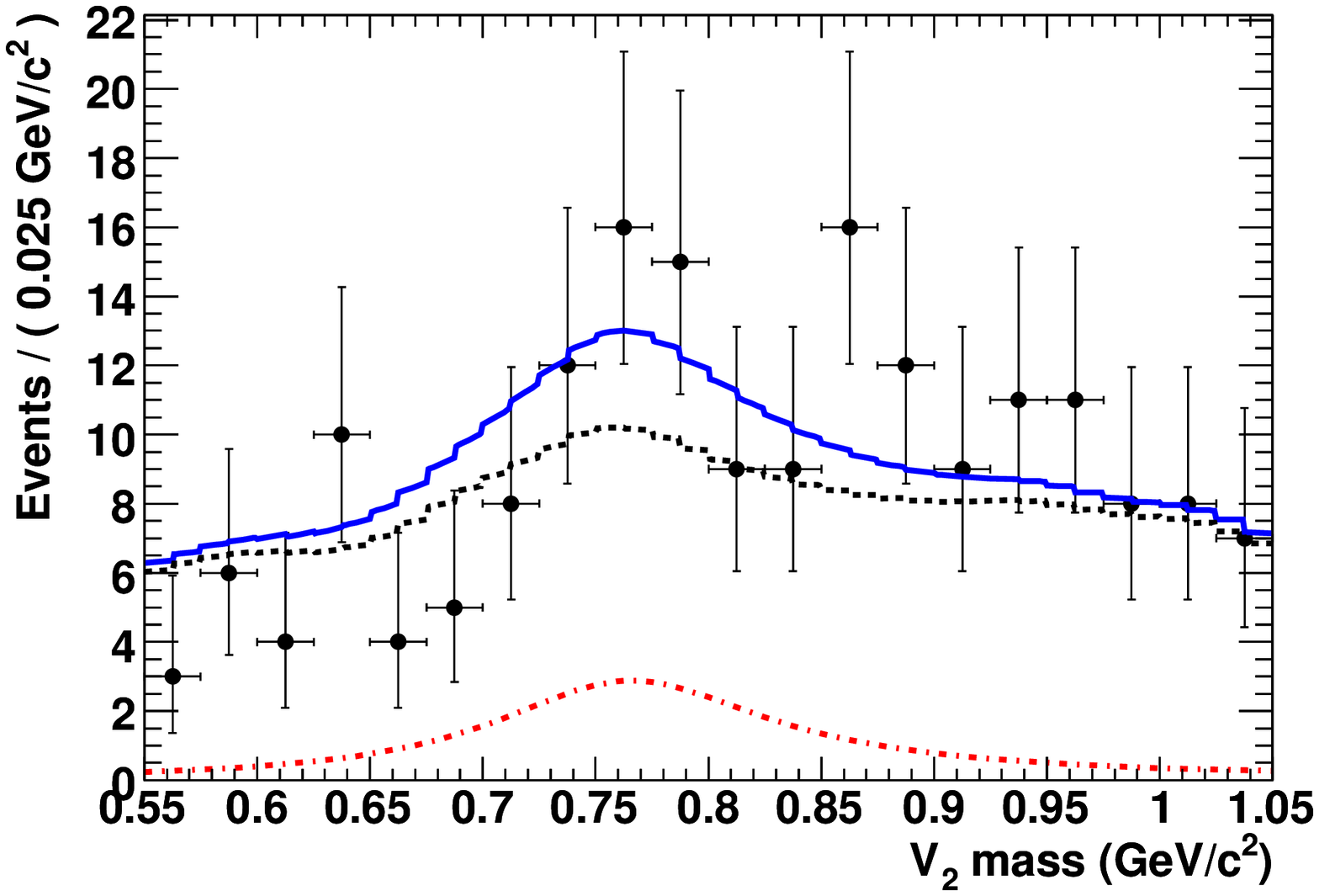}
\\
(d) & (e) \\
\includegraphics[height=1.5in]{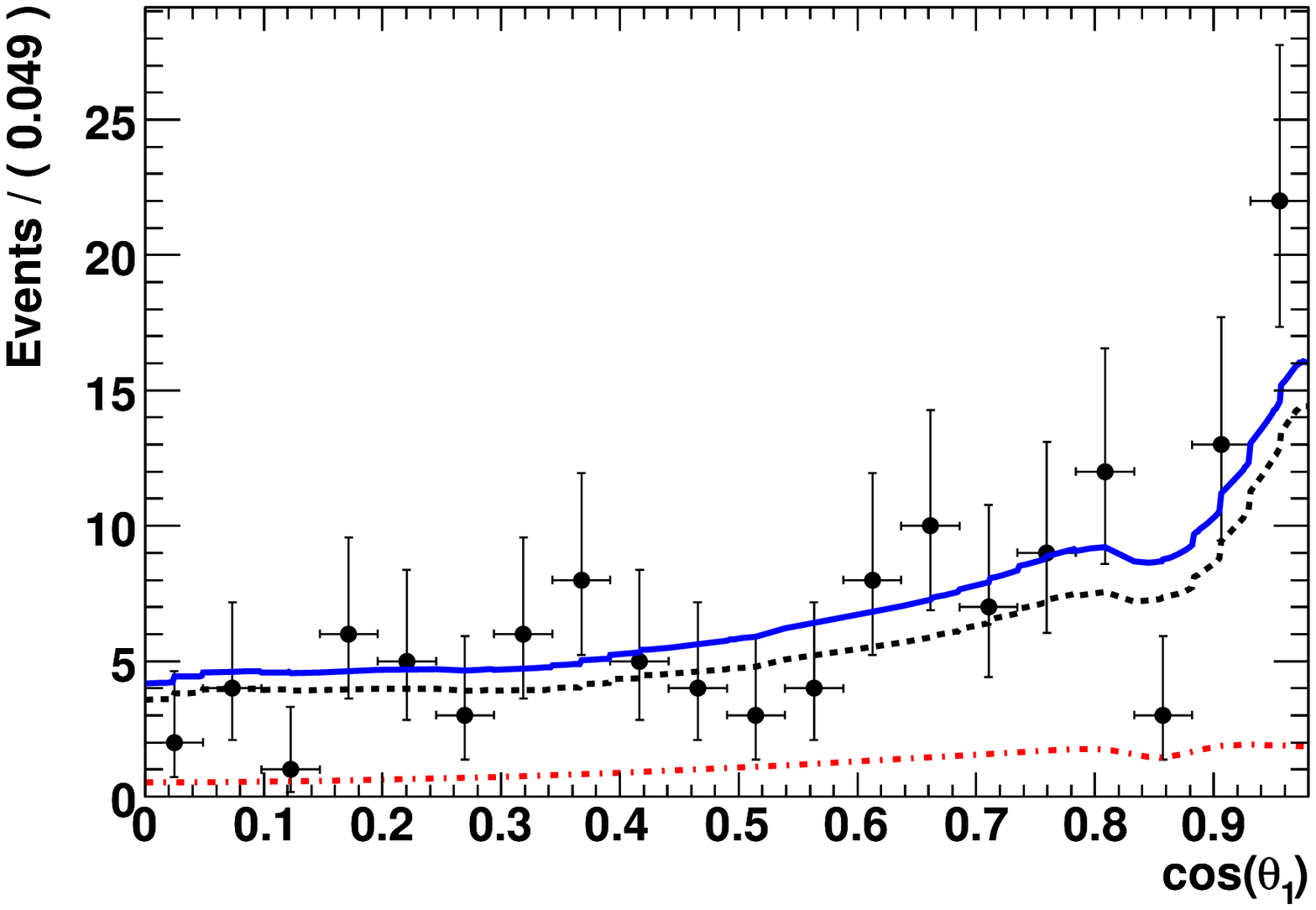}
&
\includegraphics[height=1.5in]{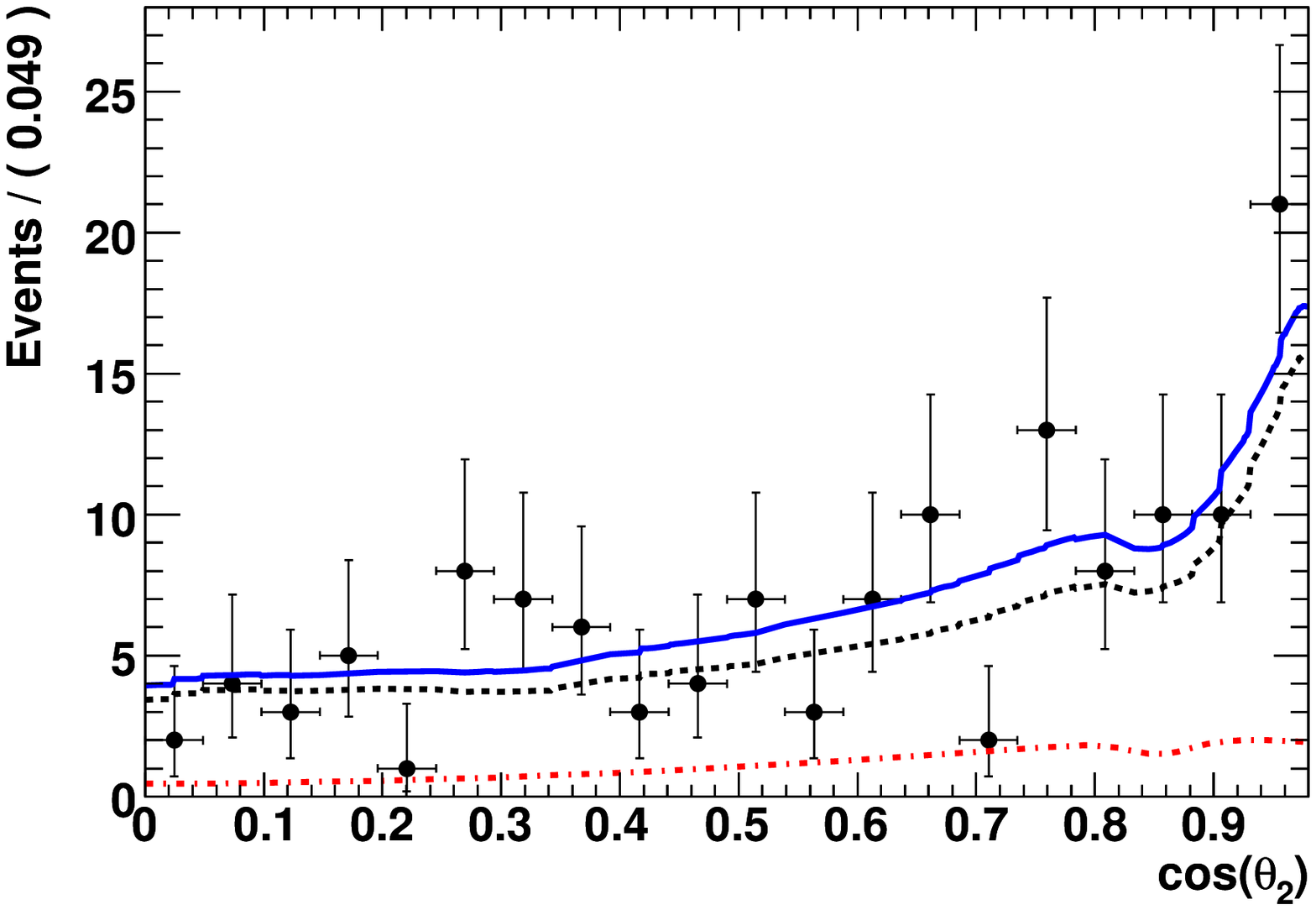}
\\
(f) & (g) \\
\end{tabular}
\end{center}
\caption{
Projections of the multidimensional fit onto
(a) $m_{\rm ES}$, (b) $\Delta E$, 
(c) event shape variable $\mathcal{E}$, 
(d,e) di-pion invariant masses $m_1$ and $m_2$, 
and (f,g) cosines of the helicity angles $\cos\theta_{1,2}$, after
a requirement on the 
signal-to-background probability ratio
with the plotted variable excluded.
This requirement enhances the fraction of
signal events in the sample while keeping approximately $25\%$ of
signal events. 
The data points are overlaid by the solid blue line,
which corresponds to the full PDF projection.
The individual $B^0\to\rho^0\rho^0$ PDF component
is also shown with a dot-dashed red line. The sum of all other PDFs
(including $B^0\to\rho^0f_0$ and $B^0\to f_0f_0$ components) is shown
as the dashed black line. 
The $D$-meson veto causes the acceptance dip seen in (f,g).
}
\label{fig:projections}
\end{figure}

\begin{figure}[f]
\begin{center}
\begin{tabular}{cc}
\includegraphics[height=1.5in]{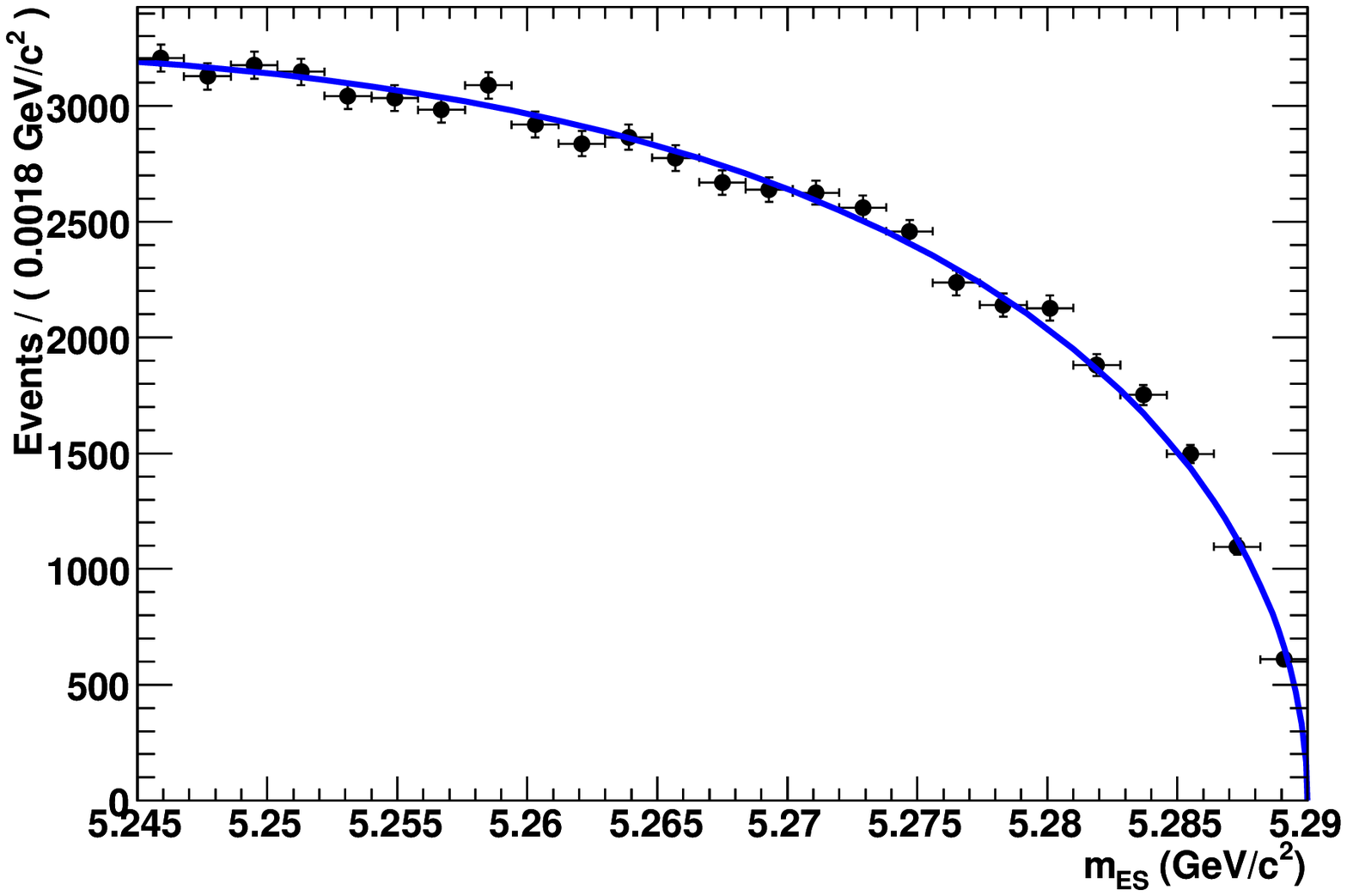}
&
\includegraphics[height=1.5in]{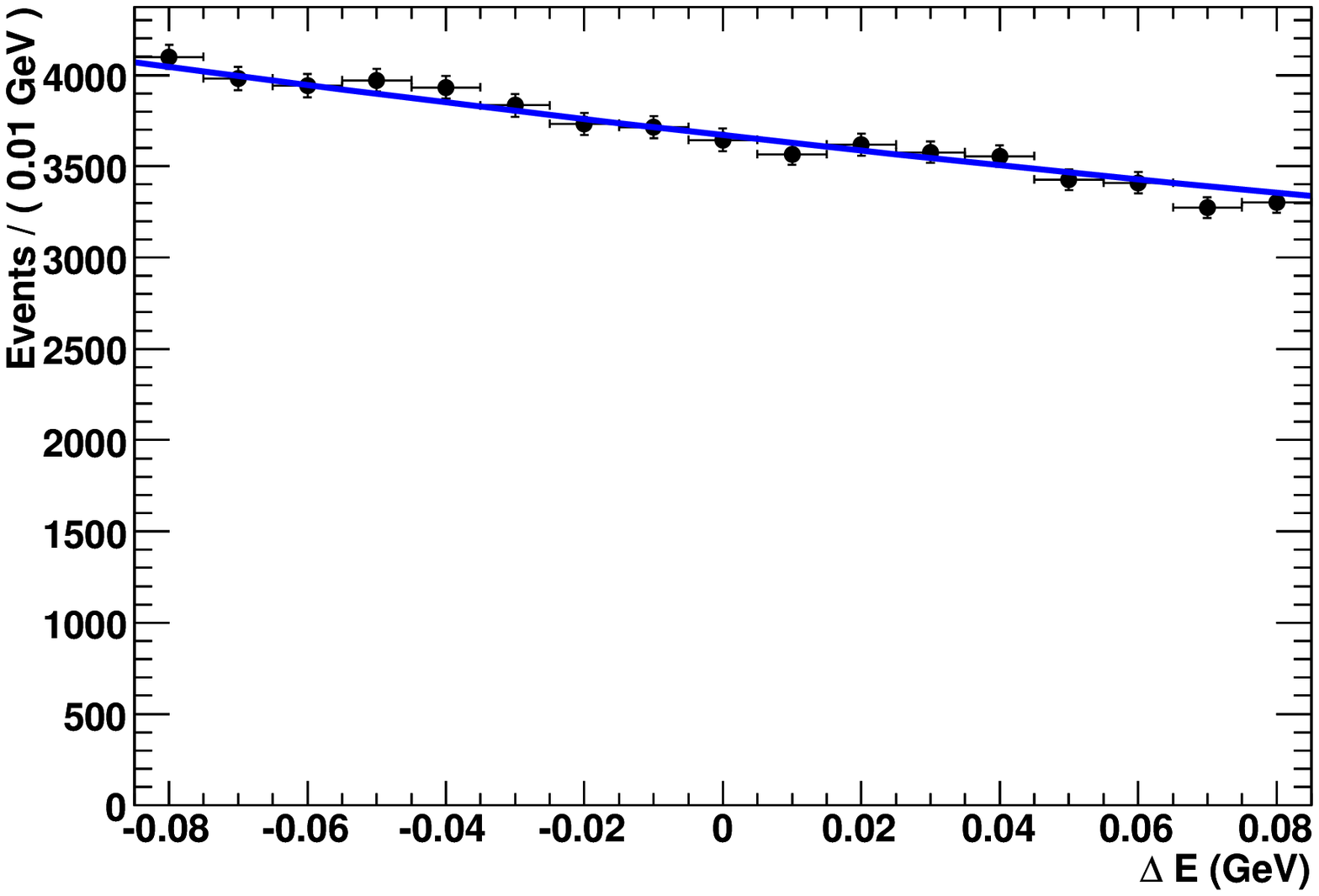}
\\
(a) & (b) \\
\includegraphics[height=1.5in]{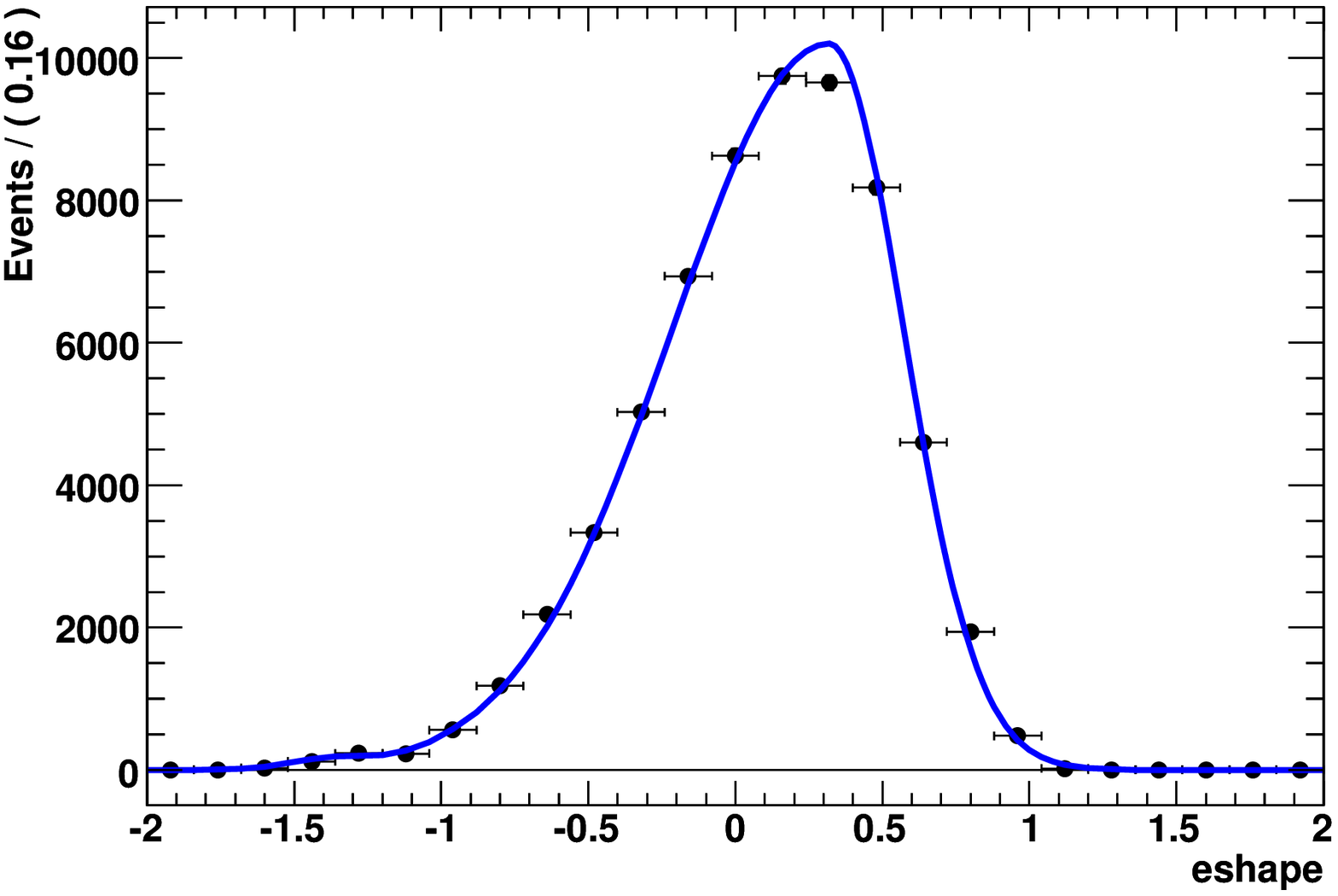}
&
\\
(c) & \\
\includegraphics[height=1.5in]{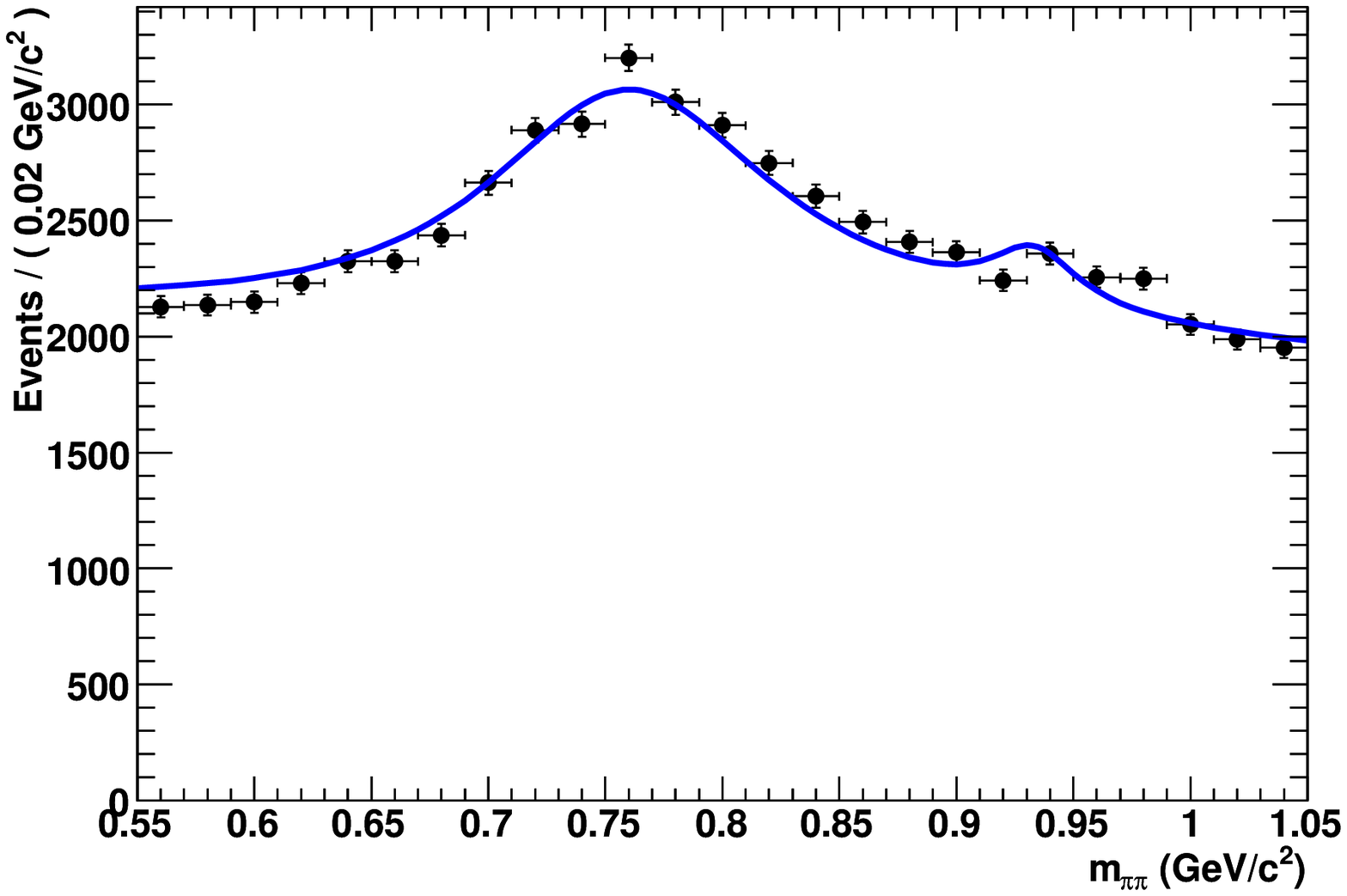}
&
\includegraphics[height=1.5in]{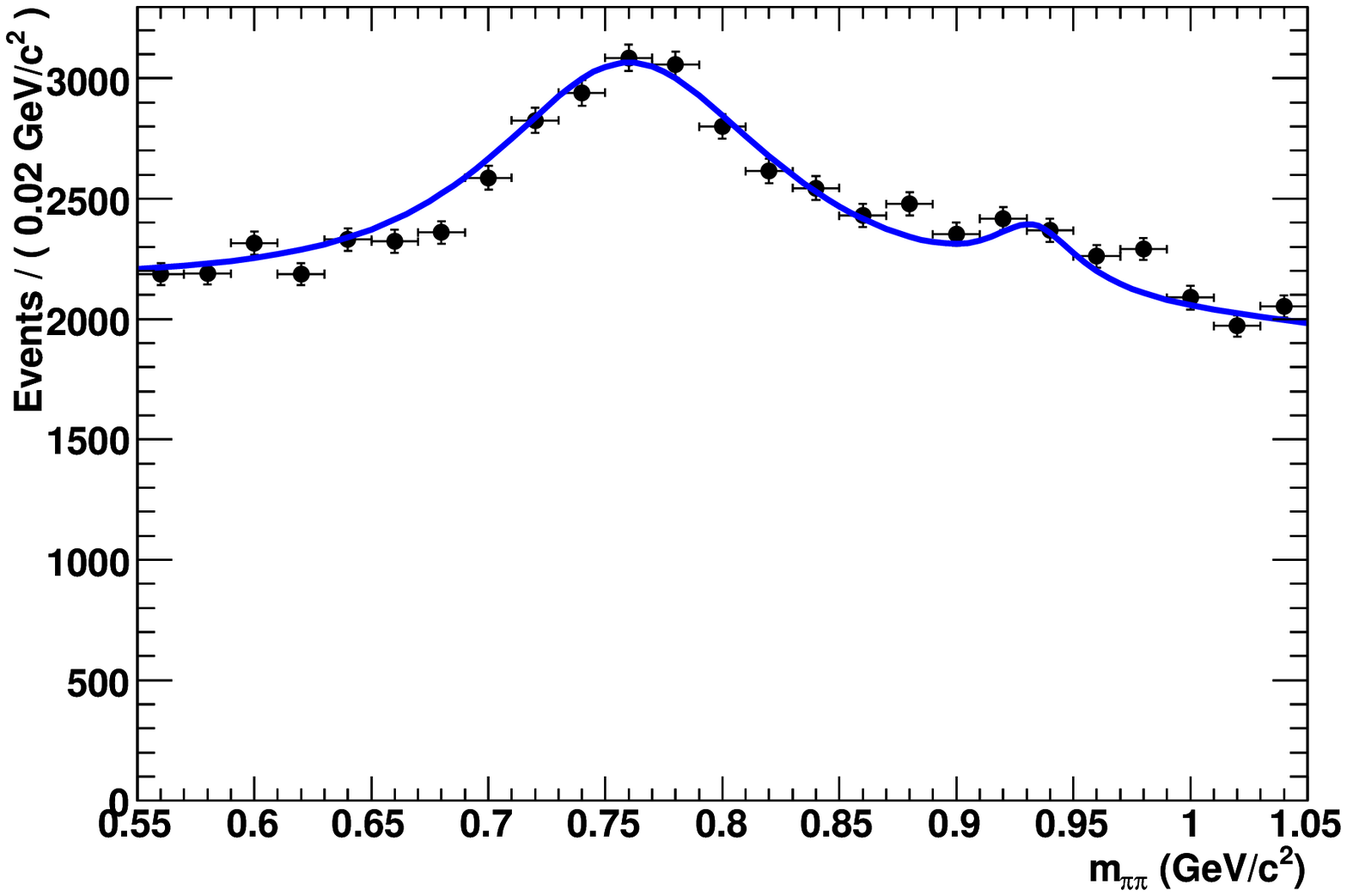}
\\
(d) & (e) \\
\includegraphics[height=1.5in]{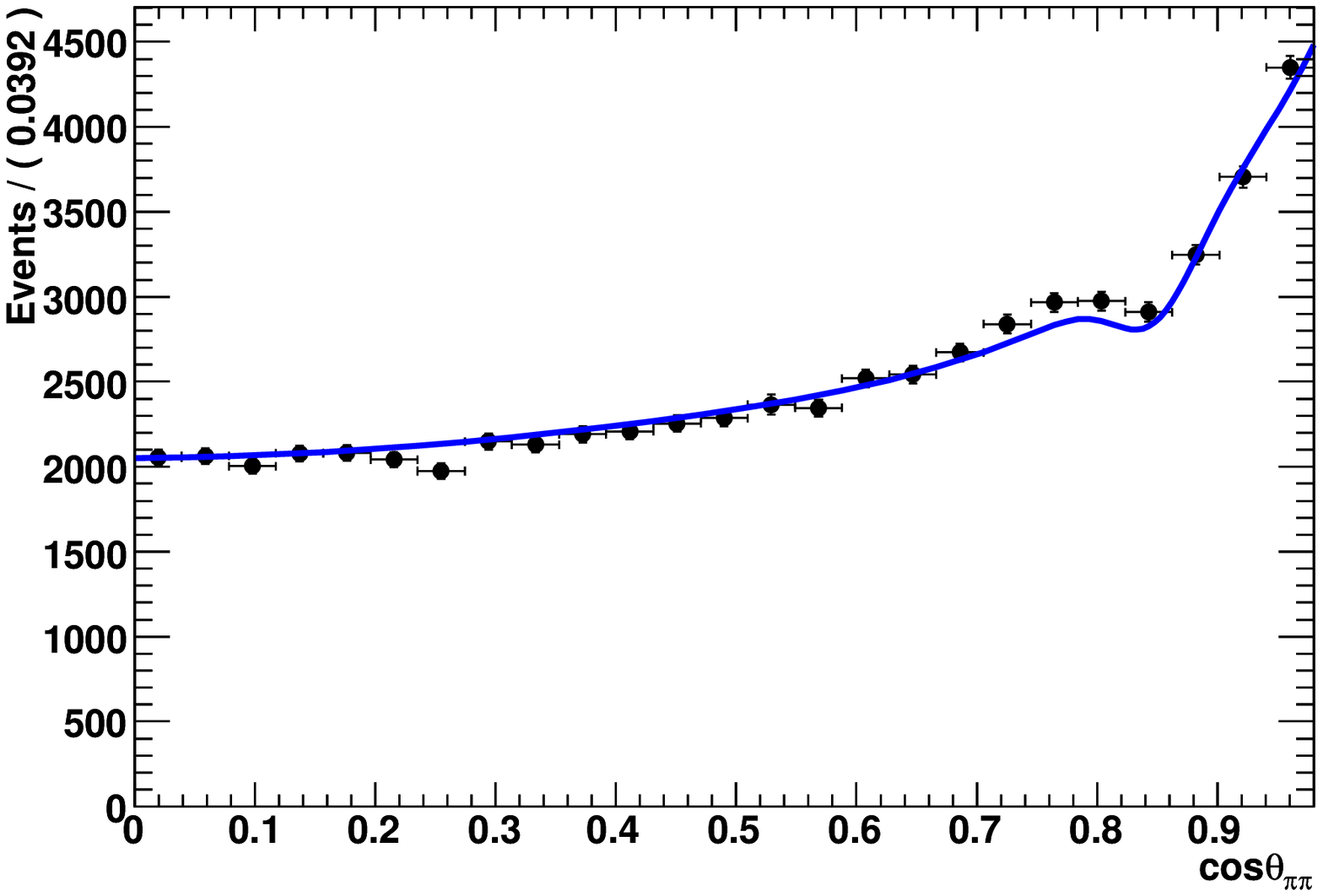}
&
\includegraphics[height=1.5in]{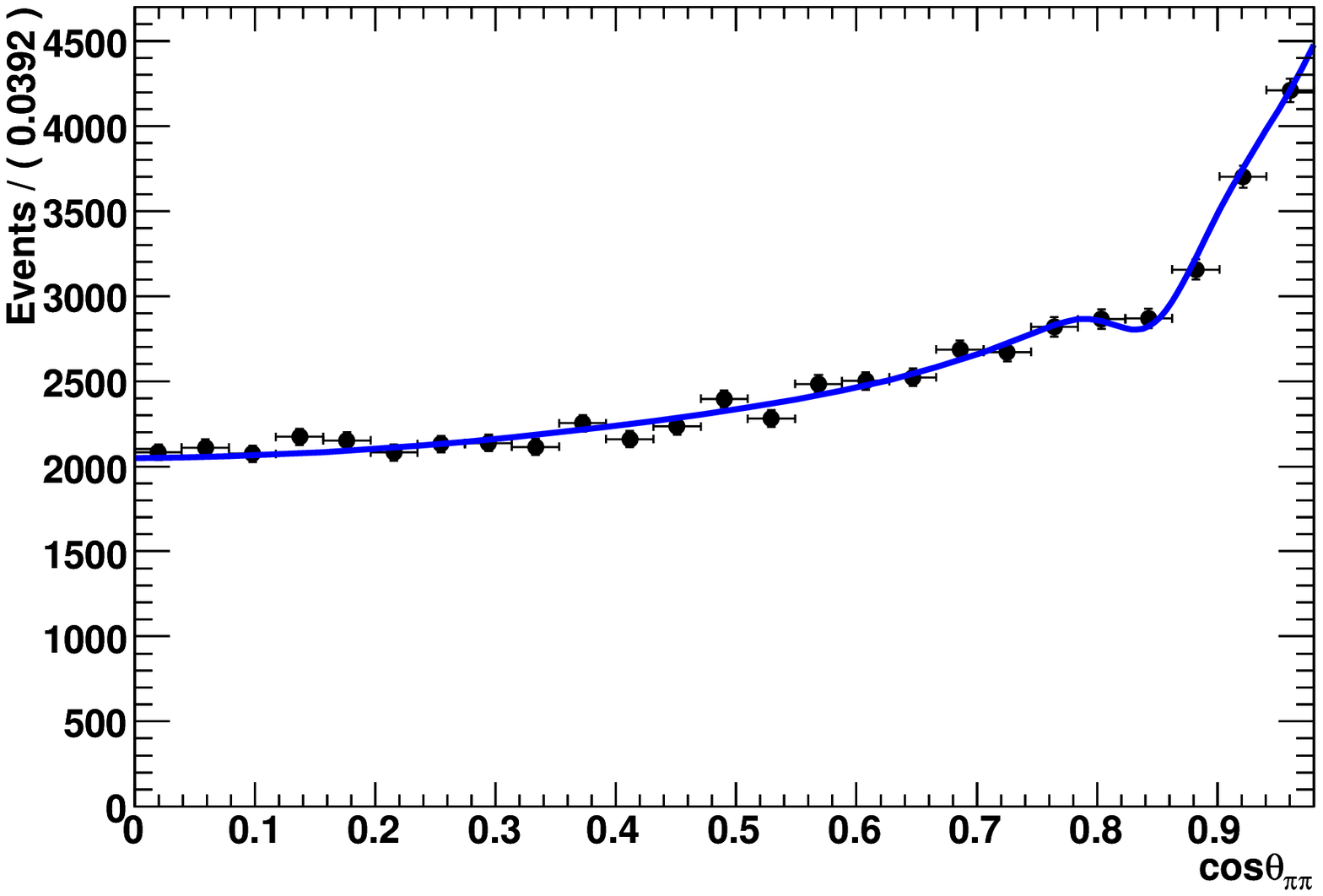}
\\
(f) & (g) \\
\end{tabular}
\end{center}
\caption{
Signal-subtracted projections (sPlots) of the multidimensional fit onto
(a) $m_{\rm ES}$, (b) $\Delta E$, 
(c) event shape variable $\mathcal{E}$, 
(d,e) di-pion invariant masses
$m_1$ and $m_2$, 
and (f,g) cosines of the helicity angles
$\cos\theta_{1,2}$. The data are weighed to enhance the continuum
$q\bar{q}$ events and effectively subtract all other fit
components. 
The data points are overlaid by the solid blue line,
which corresponds to the $q\bar{q}$ PDF component. 
}
\label{fig:splots_bkg}
\end{figure}

\begin{figure}[htb]
\begin{center}
\begin{tabular}{cc}
\includegraphics[width=3in]{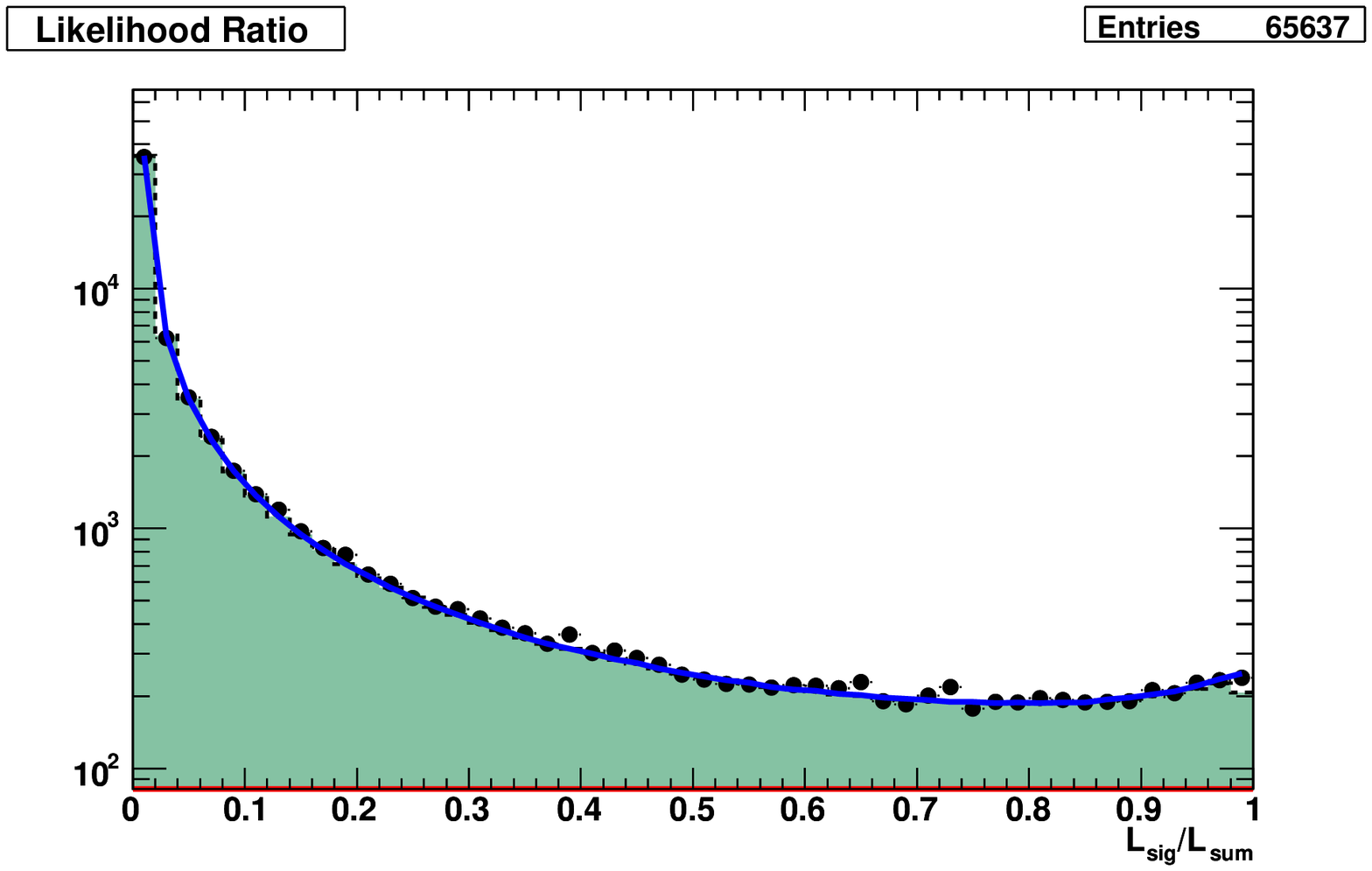} &
\includegraphics[width=3in]{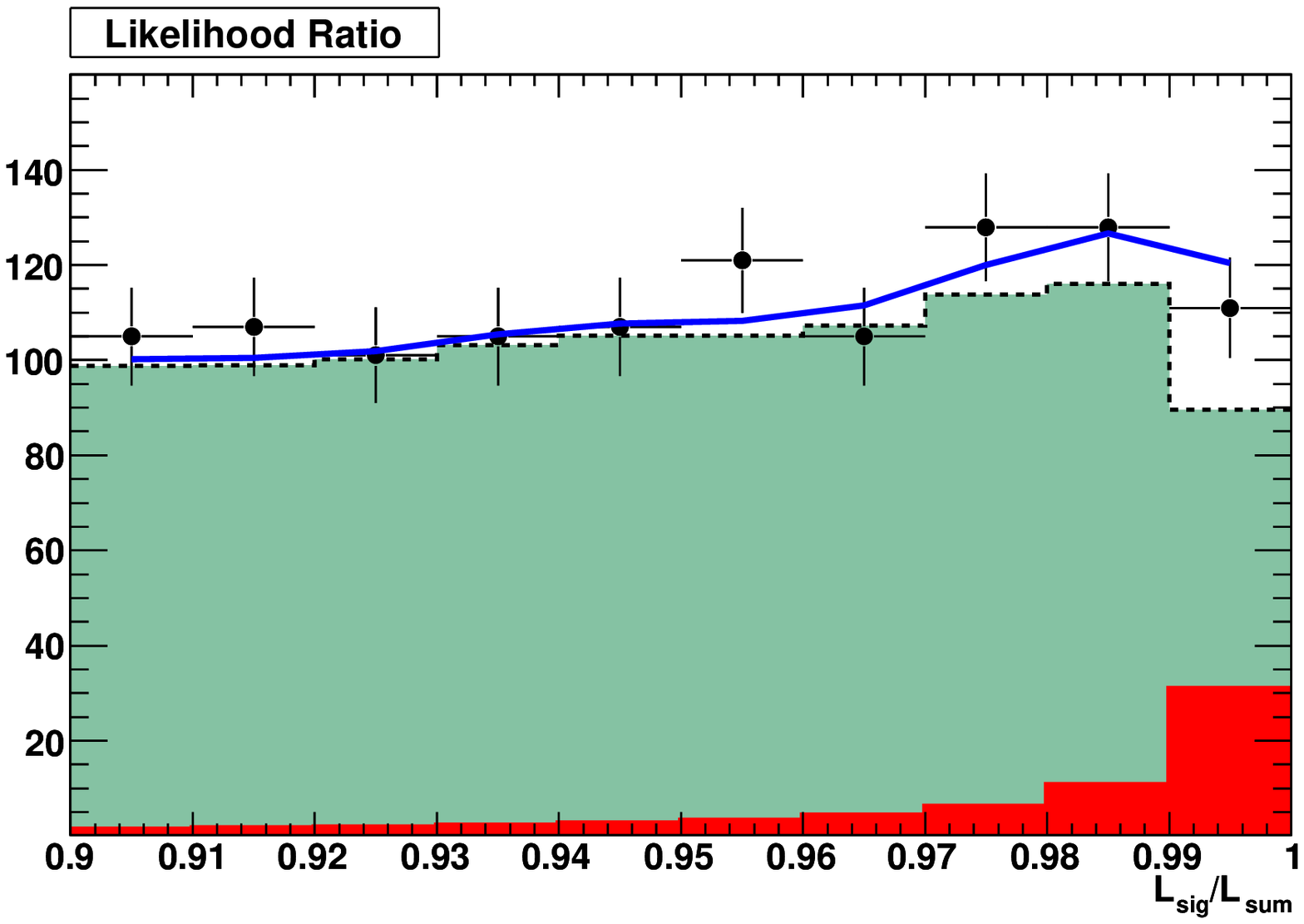} \\
(a) & (b) \\
\end{tabular}
\end{center}
\caption{\sl Likelihood ratio
$\mathcal{L}_\mathrm{sig}/\sum_i\mathcal{L}_i$, where likelihoods
$\mathcal{L}_i$ include all signal and background PDFs. Data points
are overlaid by a blue curve, which corresponds to the full
PDF. The shaded teal histogram and the black dashed line correspond to the
sum of background PDFs, and the red histogram corresponds to the signal
contribution. Full range (a) and a zoom-in into the signal region (b)
are shown.}
\label{fig:LLR}
\end{figure}

We find excellent agreement between our two approaches to modeling \B
background events. 
Table~\ref{tab:results} shows the average results, while the
difference between the two models is used to determine the
systematic uncertainty associated with \B background description.
The $\Bztorhozrhoz$ decay is observed with a significance of $3.6\sigma$,
as determined by 
the quantity $\sqrt{-2\log(\mathcal{L}_0/\mathcal{L}_{\max})}$, where 
$\mathcal{L}_{\max}$ is the maximum likelihood value, and 
$\mathcal{L}_0$ is the likelihood for a fit with the signal
contribution set to zero. Both likelihoods include systematic
uncertainties, which are assumed to be Gaussian-distributed. 
This significance level corresponds to a probability of
background fluctuation 
to the observed signal yield of $2.8\times10^{-4}$. We do not
observe significant event yields for $\Bztorhozfz$ and $\Bztofzfz$
decays, nor of the non-resonant decays $B^0\to \rho^0\pi^+\pi^-$ and
$B^0\to\pi^+\pi^-\pi^+\pi^-$. If the non-resonant contributions 
$B^0\to \rho^0\pi^+\pi^-$ and $B^0\to\pi^+\pi^-\pi^+\pi^-$ were ignored
in the fit, the significance for $\Bztorhozrhoz$ signal would go up to
$4.2\sigma$. 
Background yields are found to be
consistent with expectations. 
In Fig.~\ref{fig:projections} we show the projections of the fit results
onto $m_{\rm ES}$, $\DeltaE$, $\eshape$, $m_{1,2}$, and
$\cos\theta_{1,2}$ variables. 
Fig.~\ref{fig:splots_bkg} shows the 
distributions for the continuum $q\bar{q}$ component of the fit after
subtracting the other components~\cite{ref:sPlots}. 
Fig.~\ref{fig:LLR} shows the distribution of the likelihood ratio
$L_\mathrm{sig}/\sum_i\mathcal{L}_i$, where likelihoods 
$\mathcal{L}_i$ include all signal and background PDFs. This ratio
peaks near 1 for the signal events ( Fig.~\ref{fig:LLR}b), and is
highly peaked near zero for backgrounds.

\begin{figure}[htb]
\begin{center}
\includegraphics[height=4in]{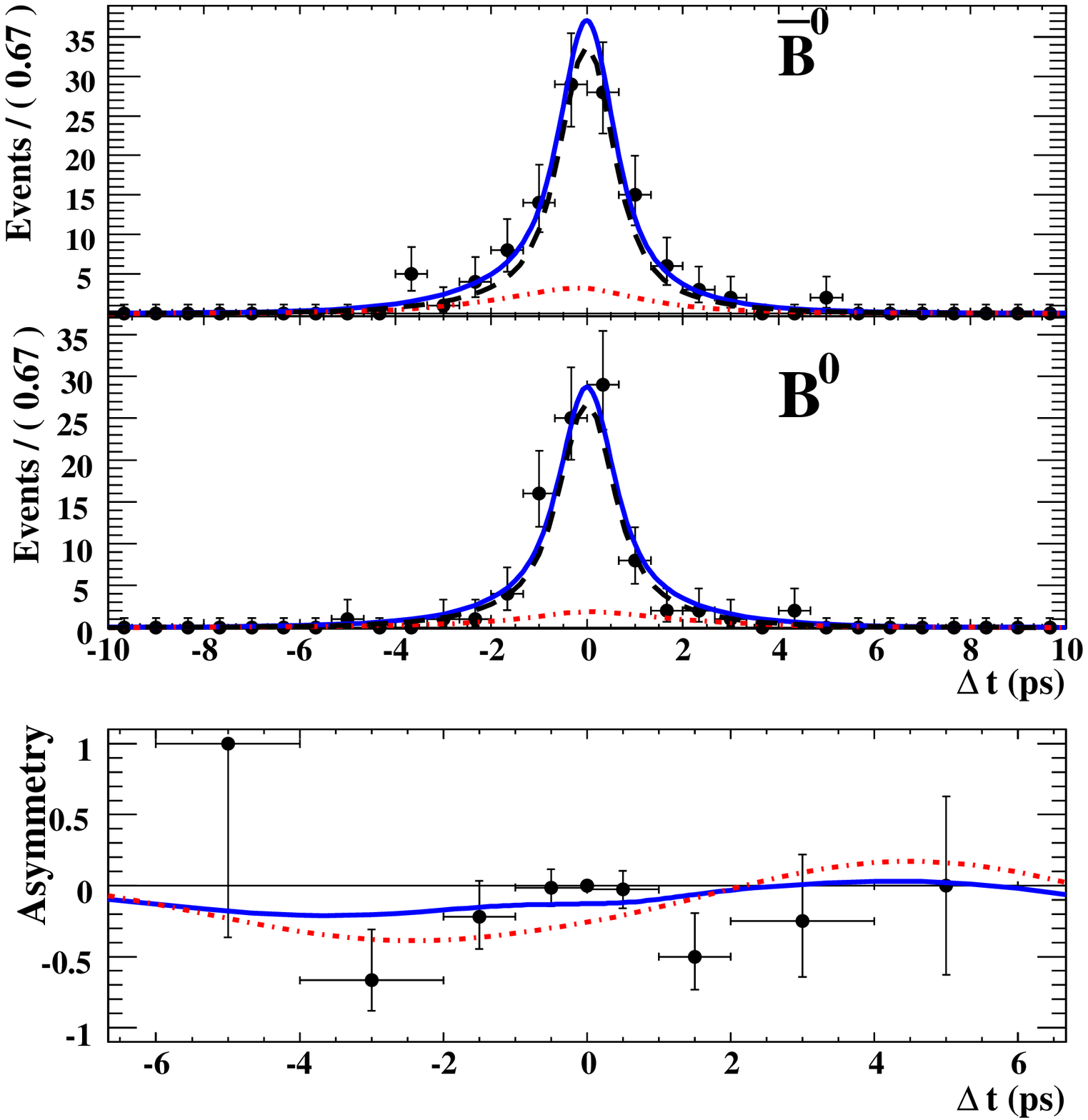} 
\end{center}
\caption{\sl Projections of the multidimensional fit onto $\Delta t$
  variable for (top) $\overline{B}^0$ tags and (middle) $B^0$
  tags. $C\! P$ asymmetry $\mathcal{A}_{C\!P}$ is shown in the bottom
  plot. The solid blue line represents the projection of the total
  PDF, the red dot-dashed line is the $B^0\to\rho^0\rho^0$
  contribution, and the dashed black line corresponds to the sum of
  all backgrounds. A likelihood cut
  $\mathcal{L}_\mathrm{sig}/\sum_i\mathcal{L}_i>0.99$ is applied to
  enhance the signal contribution.}
\label{fig:deltaT}
\end{figure}

We also fit the proper-time distribution of the data sample, and
determine the $C\! P$-violating parameters $S^{00}_L$ and $C^{00}_L$
for the longitudinal component of the $\B^0\to\rho^0\rho^0$
sample. The results are listed in Table~\ref{tab:results}. The
projection plots of $\Delta t$ distributions for $\overline{B}^0$ and
$B^0$ tags, as well as the plot of the $C\!P$ asymmetry, are shown in
Fig.~\ref{fig:deltaT}. 

\begin{table}[htb]
\caption{
Summary of results: event yields ($n$), corrected for fit bias;
fraction of longitudinal polarization ($f_L$);
selection efficiency (Eff) corresponding to measured polarization;
branching fraction (${\cal B}_{\rm sig}$),
and significance including systematic uncertainties.
The systematic errors are quoted last. 
We also show the background event yields for $a_1\pi$, $\qqbar$,
charmless, and other $\BB$ components (statistical uncertainties only). 
}
\vspace{0.2cm}
\begin{center}
\begin{tabular}{lc}
\hline\hline
                              &   \vspace*{-0.35cm} \\
Quantity                      &  Value             \\
                              &   \vspace*{-0.35cm} \\
\hline
                              &   \vspace*{-0.3cm} \\
$n$($B^0\to\rho^0\rho^0$)
                              &   $85\pm 28\pm 17$ \\
                              &   \vspace*{-0.35cm} \\
$f_L$ 
                              &   $0.70\pm 0.14\pm 0.05$ \\
                              &   \vspace*{-0.35cm} \\
Eff (\%)                      &   $23.8\pm 1.0$       \\
                              &   \vspace*{-0.35cm} \\
${\cal B}_{\rm sig}$ $(\times 10^{-6})$  &   $0.84\pm 0.29\pm 0.17$ \\
                              &    \vspace*{-0.35cm} \\
Significance, stat. only ($\sigma$)       &    $4.0$   \\
Significance, syst. included ($\sigma$)       &    $3.6$ \\
                              &    \vspace*{-0.35cm} \\
\hline
                              &    \vspace*{-0.3cm} \\
$C^{00}_L$
                              &    $0.4\pm0.9\pm0.2$ \\
                              &    \vspace*{-0.3cm} \\
$S^{00}_L$
                              &    $0.5\pm0.9\pm0.2$ \\
                              &    \vspace*{-0.35cm} \\
\hline
                              &    \vspace*{-0.3cm} \\
$n(B^0\to\rho^0f_0(980))$       &  $-11\pm 16$        \\
                              &   \vspace*{-0.3cm} \\
$n(B^0\to f_0(980) f_0(980))$ &  $6\pm 6$        \\
                              &   \vspace*{-0.3cm} \\
$n(B^0\to a_1^\pm\pi^\mp)$    & $296\pm 42$ \\
                              &   \vspace*{-0.35cm} \\
$n({\rm charmless})$          & $348\pm 64$   \\
                              &   \vspace*{-0.35cm} \\
$n({\BB})$                     & $2614\pm 134$   \\
                              &   \vspace*{-0.35cm} \\
$n({\qqbar})$                 & $62298\pm 268$  \\
                              &   \vspace*{-0.35cm} \\
\hline\hline
\end{tabular}
\end{center}
  \label{tab:results}
\end{table}

%
%


\section{SYSTEMATIC STUDIES}
\label{sec:systematics}

The systematic uncertainties for all physics parameters
are summarized in Table~\ref{tab:syst}. 
Dominant systematic uncertainties in the fit originate from 
statistical errors in the PDF parameterizations, due to the limited
number of events in the control samples, variations in the \B
background branching ratios fixed in the fit, and from the
potential fit bias. 
The PDF parameters are varied by their respective uncertainties
to derive the corresponding systematic errors. 
The fit bias is studied in a large number of Monte Carlo experiments,
in which signal and charmless \B background events are fully
simulated, and $b\to c$ background events and continuum $q\bar{q}$
events are sampled from their respective PDF. 
The uncertainty associated with the \B background model
is $4$ events for the signal yield, $0.01$ for $f_L$, $0.01$ for
$C^{00}_L$ and $0.11$ for $S^{00}_L$, derived from the difference 
between the fits for the two \B background models. 
The systematic uncertainties due to the charmless background
composition, arising from the uncertainties in the individual
branching ratios and the $C\! P$ content of \B
background~\cite{HFAG07,ref:a1piCP}, are  
$5$ events for the signal yield, $0.01$ for $f_L$, $0.18$ for
$C^{00}_L$ and $0.14$ for $S^{00}_L$.
The above systematic uncertainties do not scale with event yield
and are included in the calculation of the significance of the result.

We estimate the systematic uncertainty due to the interference 
between the $B^0\to\rho^0\rho^0$ and $a_1^{\pm}\pi^{\mp}$ decays using 
simulated samples in which the decay amplitudes for $B^0\to\rho^0\rho^0$
are generated according to this measurement
and those for $B^0\to a_1^{\pm}\pi^{\mp}$ correspond
to a branching fraction of $(33.2\pm4.8)\times 10^{-6}$~\cite{a1pi}.
Their amplitudes are modeled with a Breit-Wigner function
for all $\rho\to\pi\pi$ and $a_1\to\rho\pi$ combinations 
and their relative phase is assumed to be constant across the phase space.
The strong phases and \CP\ content of the interfering state
$a_1^{\pm}\pi^{\mp}$ are varied between zero and a maximum 
value using uniform prior distributions.
We take the RMS variation of the average signal yield
(14 events for the $\rho^0\rho^0$ yield and $0.03$ for $f_L$) 
as a systematic uncertainty.

Uncertainties in the reconstruction efficiency
arise from track finding and particle identification, and are
determined by dedicated studies on copious control data control
samples. Uncertainties due to other selection requirements,
such as vertex probability, track multiplicity, and thrust angle,
amount to $2.4\%$ for the event yields, and are negligible for the
polarization and $C\! P$ observables. 

\begin{table}[htb]
\begin{center}
\caption{\sl Summary of systematic uncertainties.}
\vspace{0.3cm}
\begin{tabular}{lccccc}
\hline\hline
Source                          &  \multicolumn{2}{c}{$n(B\to\rho^0\rho^0)$}  
                                &  $f_L$ 
                                &  $S^{00}_L$ 
                                &  $C^{00}_L$ \\
                                &  fraction & events &  & & \\
\hline
                                  \multicolumn{6}{c}{Multiplicative} \\
\hline
Number of \B mesons             &  1.1\%  &  --   &  --   &  --   &  --  \\
Event selection                 &  2.4\%  &  --   &  --   &  --   &  --  \\
PID selection                   &  2.0\%  &  --   &  --   &  --   &  --  \\
Track finding                   &  1.4\%  &  --   &  --   &  --   &  --  \\
MC statistics                   &  $<$1\% &  --   & $<0.01$ & $<0.01$ &  $<0.01$  \\
$\mathrm{a}_1\pi$ interference  &   --    & $14$ & $0.025$ & $0.07$ & $0.07$ \\
\hline
                                  \multicolumn{6}{c}{Additive} \\
\hline
PDF variation                 &  --     &  $6$ &  $0.035$ & $0.07$ & $0.11$ \\
Fit bias                      &  --     &  $4$ &  $0.014$ & $0.11$ & $0.09$ \\
\B background BR\& CP         &  --     &  $5$ &  $0.010$ & $0.14$ & $0.18$ \\
\B background model           &  --     &  $3$ &  $0.005$ & $0.11$ & $0.02$ \\     
\hline
                              &         &    &         &        & \vspace*{-0.4cm}  \\
Total                         & $3.6\%$ &  $17$ &  $0.047$ & $0.23$ & $0.24$ \\     
\hline\hline
\end{tabular}
\label{tab:syst}
\end{center}
\end{table}


\section{IMPLICATIONS FOR CKM ANGLE $\alpha$}
\label{sec:alpha}

To constrain the penguin contributions to $B\to\rho\rho$ decays, we
perform an isospin analysis,
by minimizing a $\chi^2$ term that includes the measured quantities
expressed as the lengths of the sides of the isospin triangles.
We use the measured branching fractions and
fractions of longitudinal polarization of the
$\Bptorhozrrhop$~\cite{rho0rhop2}
decays,  the measured branching fractions, polarization, and 
 \CP\ parameters $S^{+-}_{L}$ and $C^{+-}_{L}$
determined from the time evolution of the longitudinally
polarized $\Bztorhoprhom$ decay~\cite{rhoprhom}, 
and finally the branching fraction, the polarization,
and the \CP\ parameters $S^{00}_{L}$ and $C^{00}_{L}$
of $\Bztorhozrhoz$ from this analysis.
We assume uncertainties to be Gaussian and 
neglect $I=1$ isospin contributions, electroweak loop amplitudes, 
non-resonant, and isospin-breaking effects.

\begin{figure}[htb]
\begin{center}
\setlength{\epsfxsize}{0.9\linewidth}
\leavevmode\epsfbox{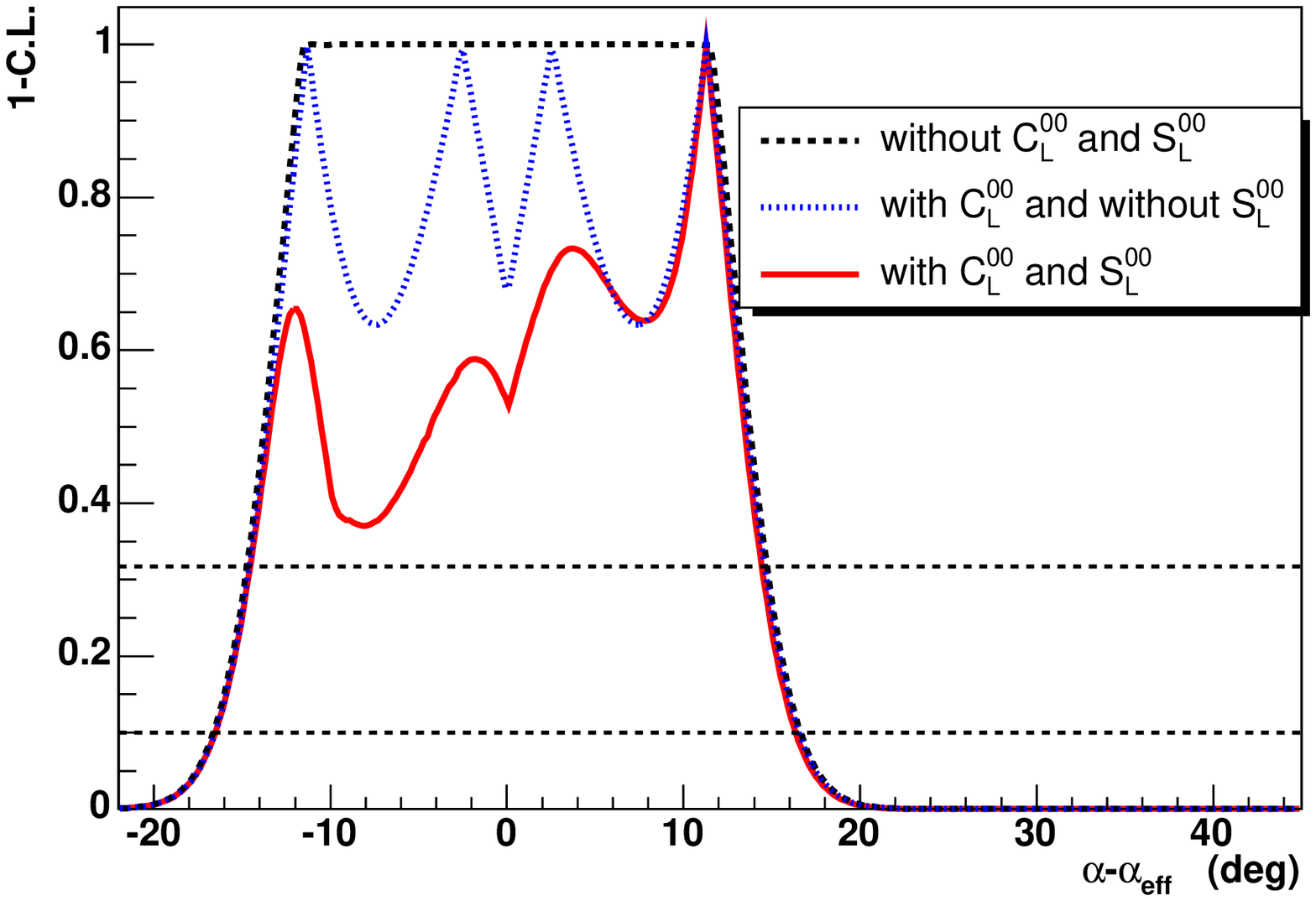}
\caption{Confidence level on $\alpha-\alpha_{\rm eff}$ obtained from 
the isospin analysis discussed in the text. The solid line CL
includes the \CP\ parameters $C^{00}_L$ and $S^{00}_L$ in the fit. 
The dotted line corresponds to the usual isospin analysis without
$S^{00}_L$. The dashed curve is obtained without the two \CP\
parameters $S^{00}_{L}$ and $C^{00}_{L}$. 
The horizontal dashed lines correspond to the $68\%$ (top) and $90\%$ 
(bottom) CL  intervals.}
\label{fig:DeltaAlphaScan}
\end{center}
\end{figure}

Using the \Bztorhozrhoz measurement we 
obtain a 68\% (90\%) CL limit on
$|\alpha-\alpha_{\rm eff}|<14.5^\circ$ ($<16.5^\circ$)
where $\alpha_{\rm eff}$ is constrained by the relation
$\sin(2\alpha_{\rm eff})= S^{+-}_{L}/({1-C^{+-2}_{L}})^{1/2}$.
Fig.~\ref{fig:DeltaAlphaScan} shows the confidence level with 
or without using the measured  \CP\ parameters $S^{00}_{L}$ and $C^{00}_{L}$
in the isospin analysis fit. We observe the four solutions around zero
as in the $B\to\pi\pi$ isospin analysis, but here, thanks to the additional 
constraint on $S^{00}_L$, one of the four solutions 
$\alpha-\alpha_{\rm eff}=+11.3^{\circ}$. 
is favored. 

\section{CONCLUSION}
\label{sec:Conclusion}

In summary, we confirm our earlier evidence for \Btozz\ decays with a
$3.6\sigma$ 
significance. We measure the $\Btozz$ branching fraction of
$(0.84\pm 0.29\pm 0.17)\times 10^{-6}$
and determine the longitudinal polarization fraction for these
decays of $f_L = 0.70\pm 0.14\pm 0.05$. We also constrain the 
$CP$ parameters $C^{00}_L$ and $S^{00}_L$ for the longitudinal part
of the \Btozz\ final state:
\begin{eqnarray}
S^{00}_L &=& 0.5 \pm 0.9 \pm 0.2 \nonumber \\
C^{00}_L &=& 0.4 \pm 0.9 \pm 0.2 \nonumber 
\end{eqnarray}
where the first uncertainty is statistical and the second is
systematic. 
These measurements combined with those for
$B^+\to\rho^0\rho^+$ and $B^0\to\rho^+\rho^-$ decays provide
a constraint on the penguin uncertainty
in the determination of the CKM unitarity angle $\alpha$.
We find no significant evidence for the decays $B^0\to \rho^0f_0$,
$B^0\to f_0f_0$, $B^0\to \rho^0\pi^+\pi^-$ and
$B^0\to\pi^+\pi^-\pi^+\pi^-$.


\section{ACKNOWLEDGMENTS}
\label{sec:Acknowledgments}

We are grateful for the
extraordinary contributions of our \pep2\ colleagues in
achieving the excellent luminosity and machine conditions
that have made this work possible.
The success of this project also relies critically on the
expertise and dedication of the computing organizations that
support \babar.
The collaborating institutions wish to thank
SLAC for its support and the kind hospitality extended to them.
This work is supported by the
US Department of Energy
and National Science Foundation, the
Natural Sciences and Engineering Research Council (Canada),
the Commissariat \`a l'Energie Atomique and
Institut National de Physique Nucl\'eaire et de Physique des Particules
(France), the
Bundesministerium f\"ur Bildung und Forschung and
Deutsche Forschungsgemeinschaft
(Germany), the
Istituto Nazionale di Fisica Nucleare (Italy),
the Foundation for Fundamental Research on Matter (The Netherlands),
the Research Council of Norway, the
Ministry of Science and Technology of the Russian Federation,
Ministerio de Educaci\'on y Ciencia (Spain), and the
Science and Technology Facilities Council (United Kingdom).
Individuals have received support from
the Marie-Curie IEF program (European Union) and
the A. P. Sloan Foundation.



\begin{thebibliography}{99}
\bibitem{footnote}
Charge conjugate $B$ decay modes are implied in this paper.
\end{thebibliography}
\end{document}